\documentclass[aps,prx,twocolumn,superscriptaddress,nofootinbib]{revtex4}
\usepackage{amssymb}
\usepackage{graphicx}
\usepackage{epsf,epsfig}
\usepackage{bm}
\usepackage{bbm}
\usepackage{amsfonts}
\usepackage{amsmath}
\usepackage{graphics}
\usepackage[normalem]{ulem}
\usepackage{color}

\newcommand{\be}{\begin{eqnarray}}
\newcommand{\ee}{\end{eqnarray}}
\newcommand{\ba}{\begin{array}}
\newcommand{\ea}{\end{array}}
\newcommand{\no}{\nonumber}
\newcommand{\tr}{\mbox{tr}}
\newcommand{\Tr}{\mbox{Tr}}

\newcommand{\eps}{\varepsilon}

\newcommand{\bfr}{{\bf r}}

\newcommand{\bfp}{{\bf p}}
\newcommand{\bfq}{{\bf q}}

\newcommand{\bfs}{{\bf s}}

\newcommand{\bfsigma}{{\bm \sigma}}

\newcommand{\bftheta}{{\bm \theta}}
\newcommand{\bfTheta}{{\bm \Theta}}

\begin{document}

\title{Non-linear sigma model with particle-hole asymmetry for the disordered two-dimensional electron gas}

\author{Georg Schwiete}
\affiliation{Department of Physics and Astronomy, The University of Alabama, Tuscaloosa, Alabama 35487, USA}

\begin{abstract}

The non-linear sigma model is a well-established theoretical tool for studies of transport and thermodynamics in disordered electronic systems. The conventional sigma model approach for interacting systems does not account for particle-hole asymmetry. It is therefore not suited for studying quantities that are sensitive to this effect such as the thermoelectric transport coefficient. Here, we derive a minimal extension of the Keldysh non-linear sigma model tailored for two-dimensional interacting systems. We argue that this model can be used to systematically study the combined effect of interactions and disorder on thermoelectric transport. As a first step in this direction, we use the model to analyze the structure of the heat density-density correlation function and calculate interaction corrections to its static part. The calculation of interaction corrections to the dynamical part of the correlation function and the thermoelectric transport coefficient is left for future work.

\end{abstract}

\maketitle

\section{Introduction}

The nonlinear sigma model (NL$\sigma$M) formalism is a field theoretical approach to the description of diffusive electron dynamics. The NL$\sigma$M for interacting disordered electron systems was introduced by Finkel'stein~\cite{Finkelstein83}, building upon earlier work on non-interacting systems~\cite{Wegner79,Efetov80}. The formalism has since been used for numerous theoretical studies \cite{DSLD:Finkelstein90,Finkelstein94,Belitz94RMP,Baranov99,Kamenev99,Chamon99,Punnoose05,Pruisken07,Levchenko07,Kamenev11,Koenig15,Burmistrov16,Schwiete14,Schwiete14a,Schwiete14b,Schwiete16a,Schwiete16b,Liao17,Liao18}. The Finkel'stein model contains a small number of parameters: the diffusion, frequency and interaction constants. These parameters characterize the diffusive motion of electrons and are closely related to transport coefficients and thermodynamic quantities such as the the conductivity, spin susceptibility or the specific heat \cite{Finkelstein83,Finkelstein84,Castellani86}. At low temperatures, these quantities acquire logarithmic corrections in two-dimensional systems~\cite{Altshuler85,Lee85,DSLD:Finkelstein90,Belitz94RMP,DiCastro04,Finkelstein10}, which can be computed efficiently by means of a renormalization group (RG) analysis~\cite{Finkelstein83,Castellani84,Castellani87,Fabrizio91}. 

In recent years, it has become clear that unlike for the electric conductivity, not all logarithmic corrections to the thermal conductivity are of the RG type. For a thorough analysis of this problem, the NL$\sigma$M formalism was generalized to thermal transport studies in Refs.~\cite{Schwiete14a,Schwiete14b,Schwiete16a}. These studies confirmed the results of a diagrammatic RG analysis~\cite{Castellani87}, and also verified the existence of additional logarithmic corrections to the thermal conductivity~\cite{Livanov91,Raimondi04,Niven05,Catelani05,Michaeli09}. The latter  corrections appear for systems with long-range Coulomb interactions and arise from electronic energies that are lower than those relevant for the RG corrections. In Ref.~\cite{Schwiete16b}, the RG results for thermal transport were be merged with the corrections originating from low energies. This step was crucial for finding the thermal conductivity at low temperatures and for analyzing the resulting violation of the Wiedemann-Franz law. In view of these developments, it would be desirable to adapt the NL$\sigma$M formalism to the analysis of thermoelectric transport phenomena as well. Thermoelectric transport, unlike electric and thermal transport, is very sensitive to deviations from particle-hole symmetry. Indeed, in a perfectly particle-hole symmetric system, the thermoelectric transport coefficient vanishes. 
As a consequence, theoretical studies of this coefficient require a higher accuracy compared to electric and thermal transport, and the conventional NL$\sigma$M~\cite{Finkelstein83,Schwiete14b} is not suited for this purpose. In order to overcome this limitation, we introduce here a minimal extension of the conventional Finkel'stein model. The model is specifically tailored for two-dimensional systems with quadratic dispersion. We argue that this model can, for example, be used for a comprehensive study of logarithmic corrections to the thermoelectric transport coefficient in the two-dimensional disordered electron gas. The generalized NL$\sigma$M reflects the particle-hole asymmetry of the underlying microscopic model by accounting for energy-dependent deviations of the electron velocity from the Fermi velocity. Specifically, the non-constancy of the velocity manifests itself in the form of a frequency dependence of the diffusion coefficient that is absent in the conventional NL$\sigma$M approach. We find that the generalized model can be obtained from the conventional Finkel'stein model by (i) replacing the $\hat{Q}$ field by $\hat{Q}+\frac{1}{4i}D_\eps'(\nabla \hat{Q})^2$, where $D_\eps'$ is the derivative of the diffusion coefficient with respect to frequency, and (ii) including a separate contribution to the resulting four-gradient term. The contribution with four gradients in (ii) is a result of a non-trivial integration over massive modes in the derivation of the NL$\sigma$M, a mechanism previously discussed in Ref.~\cite{Wang94}. 

We include two types of source fields into the derivation of the generalized NL$\sigma$M action: a scalar potential coupling to the density and a gravitational potential coupling to the heat density. These source fields can be used for obtaining heat and charge densities and the heat density-density correlation function from the NL$\sigma$M. Knowledge of this correlation function is sufficient for finding the thermoelectric transport coefficient \cite{Fabrizio91}. For the sake of simplicity, we consider a model with Fermi-liquid type short range interactions. The resulting NL$\sigma$M is a generalization of the model derived for the calculation of the heat density-heat density correlation function in Refs.~\cite{Schwiete14a,Schwiete14b}. In analogy to these works, we employ the Keldysh real-time formalism \cite{Schwinger61,Kadanoff62,Keldish65,Kamenev11}. We analyze the structure of the heat density-density correlation function in the absence of interaction corrections with the help of the generalized NL$\sigma$M and make contact with the result obtained from conventional Boltzmann transport theory \cite{Ziman01}. As a first application to the calculation of interaction corrections, we study the static part of the heat density-density correlation function. Since the generalized NL$\sigma$M in the presence of source fields has a rather non-trivial structure, we perform the calculation in two different ways, thereby testing different terms in the model. We demonstrate that both routes lead to the same results. The results are also consistent with the diagrammatic analysis of Ref.~\cite{Fabrizio91}. Unlike the density of states in the clean case, which is constant, the disorder averaged density of states in two-dimensional systems has a weak energy dependence. In order to explore the implications of this observation we study how the result for the heat density-density correlation function in the ladder approximation is modified when this energy dependence is taken into account. We argue, that a further generalization of the NL$\sigma$M that includes a non-constant density of states is not required for studies of the leading interaction corrections in two-dimensional systems. 

This paper is structured as follows. In Sec.~\ref{sec:hdcf}, we discuss the structure of the heat density-density correlation function in the ladder approximation. We analyze the role of the frequency dependence of the diffusion coefficient and of the density of states. In Sec.~\ref{sec:genNLSM}, we introduce the generalized NL$\sigma$M with particle-hole asymmetry and source fields, discuss its symmetry properties, and use this formalism to reproduce the ladder approximation for the heat density-density correlation function in the constant density of states approximation. Sec.~\ref{sec:Heatdens} is devoted to the calculation of interaction corrections to the static part of the heat density-density correlation function. In Sec.~\ref{sec:NLSMderivation}, we present the derivation of the generalized Keldysh NL$\sigma$M introduced in Sec.~\ref{sec:genNLSM}. We first restrict ourselves to non-interacting systems. This allows us to stress the key points in a simplified set-up. We then generalize the derivation to include electron-electron interactions as well as source fields. In Sec.~\ref{sec:nonconstant}, we address the issue of the weak energy dependence of the disorder-averaged density of states. We conclude in Sec.~\ref{sec:conclusion}.

\section{The heat density-density correlation function}
\label{sec:hdcf}

\subsection{General structure}
In linear response, the thermoelectric transport coefficient can be obtained from the heat current-current correlation function. An alternative route proceeds via the heat density-density correlation function. Here, we will make use of the second possibility and develop a formalism that allows us to study to the retarded heat density-density correlation function 
\begin{align}
\chi_{kn}(x_1,x_2)=-i\theta(t_1-t_2)\langle[\hat{k}(x_1),\hat{n}(x_2)]\rangle_T,\label{eq:chikn}
\end{align}
as well as the closely related density-heat density correlation function $\chi_{nk}(x_1,x_2)=-i\theta(t_1-t_2)\langle[\hat{n}(x_1),\hat{k}(x_2)]\rangle_T$. In Eq.~\eqref{eq:chikn}, $\hat{k}=\hat{h}-\mu\hat{n}$ is the heat density operator, where $\mu$ is the chemical potential, $\hat{n}$ is the density operator, $x=(\bfr,t)$ comprises spatial coordinates $\bfr$ and time $t$, and the angular brackets denote thermal averaging. As usual, the averaging over disorder configurations establishes translational invariance: $\langle\chi_{kn}(x_1,x_2)\rangle_{dis}=\chi_{kn}(x_{1}-x_2)$. After a Fourier transformation, the correlation function is expected to take the following form in the diffusive limit \cite{Fabrizio91}
\begin{align}
\chi_{kn}(\bfq,\omega)&=\frac{D_{n} \bfq^2 D_k \bfq^2 \chi^{st}_{kn}+i L\bfq^2 \omega}{(D_{n}\bfq^2-i\omega)(D_k\bfq^2-i\omega)}.\label{eq:chikngeneral}
\end{align}
Here, $D_{n}$ and $D_k$ are the diffusion coefficients for charge and heat transport. These coefficients also enter the heat density-heat density correlation function, $\chi_{kk}(\bfq,\omega)=-TcD_k\bfq^2/(D_k\bfq^2-i\omega)$, where $c$ is the specific heat, and the density-density correlation function $\chi_{nn}(\bfq,\omega)=-\partial_\mu n D_{n}\bfq^2/(D_{n}\bfq^2-i\omega)$. In Eq.~\eqref{eq:chikngeneral}, $L$ is the thermoelectric transport coefficient. This coefficient is related to the Seebeck coefficient $S$ as $S=eL/(\sigma T)$, where $e$ is the charge of the electron and $\sigma$ is the electric conductivity. The correlation function is very sensitive to the order of the two limits $\bfq\rightarrow 0$ and $\omega\rightarrow 0$, 
\begin{align}
\chi_{kn}(\bfq=0,\omega\rightarrow 0)&=0\label{eq:cons}\\
\chi_{kn}(\bfq\rightarrow 0,\omega=0)&\equiv\chi_{kn}^{st}=-T\partial_T \left\langle \hat{n}\right\rangle_T.\label{eq:susc}
\end{align}
Eq.~\eqref{eq:cons} reflects the conservation laws of energy and particle number, while Eq.~\eqref{eq:susc} relates the static part of the correlation function to a thermodynamic susceptibility. When $k$ and $n$ are both even under time reversal, $\chi_{kn}$ and $\chi_{nk}$ are closely related, $\chi_{kn}(\bfr,\bfr',\omega)=\chi_{nk}(\bfr',\bfr,\omega)$~\cite{Kadanoff63}. As a consequence, the equality $\chi_{kn}(\bfq,\omega)=\chi_{nk}(\bfq,\omega)$ holds in the diffusive limit.

\subsection{Ladder approximation} 
\label{sec:ladder}

For the sake of clarity, we restrict our study to a model Hamiltonian with short-range interactions, as well as quadratic dispersion and a white noise disorder potential (for details, see Sec.~\ref{sec:model}). Interactions are characterized by the Fermi liquid parameters $F_0^{\rho,\sigma}$ for the singlet and triplet channels. In this section, we discuss the heat density-density correlation function $\chi_{kn}$ in the ladder approximation. In this approximation, interaction corrections resulting from loop integrations over small momenta and frequencies of diffusion modes are neglected. The function $\chi_{nk}$ can be treated in analogy. It is useful to present $\chi_{kn}$ as the sum of static and dynamical parts as 
\begin{align}
\chi_{kn}(\bfq,\omega)&=\chi_{kn}^{st}+\chi_{kn}^{dyn}(\bfq,\omega),
\end{align}
where the dynamical part is denoted as $\chi^{dyn}_{kn}(\bfq,\omega)$. In the absence of interaction corrections, the static part of the correlation function is given as 
\begin{align}
\chi_{kn}^{st,0}=-z_1^0T\partial_Tn_0=-z_1^0T \overline{c}_{0,\eps}'\label{eq:chst}.
\end{align}
In this relation, $n_0$ denotes the density in the absence of interaction corrections, the frequency-dependent specific heat $\bar{c}_{0,\eps}=2\pi^2 T\bar{\nu}_\eps/3$ is related to the (disorder averaged) density of states $\bar{\nu}_\eps$, and $z_1^0=(1+F_0^\rho)^{-1}$ is the Fermi-liquid renormalization of the density vertex.  Here and below, we denote the derivative with respect to the frequency $\eps$ by a prime, $f'_\eps=\partial_\eps f_\eps|_{\eps=0}$.

\begin{figure}[tb]
\includegraphics[width=6cm]{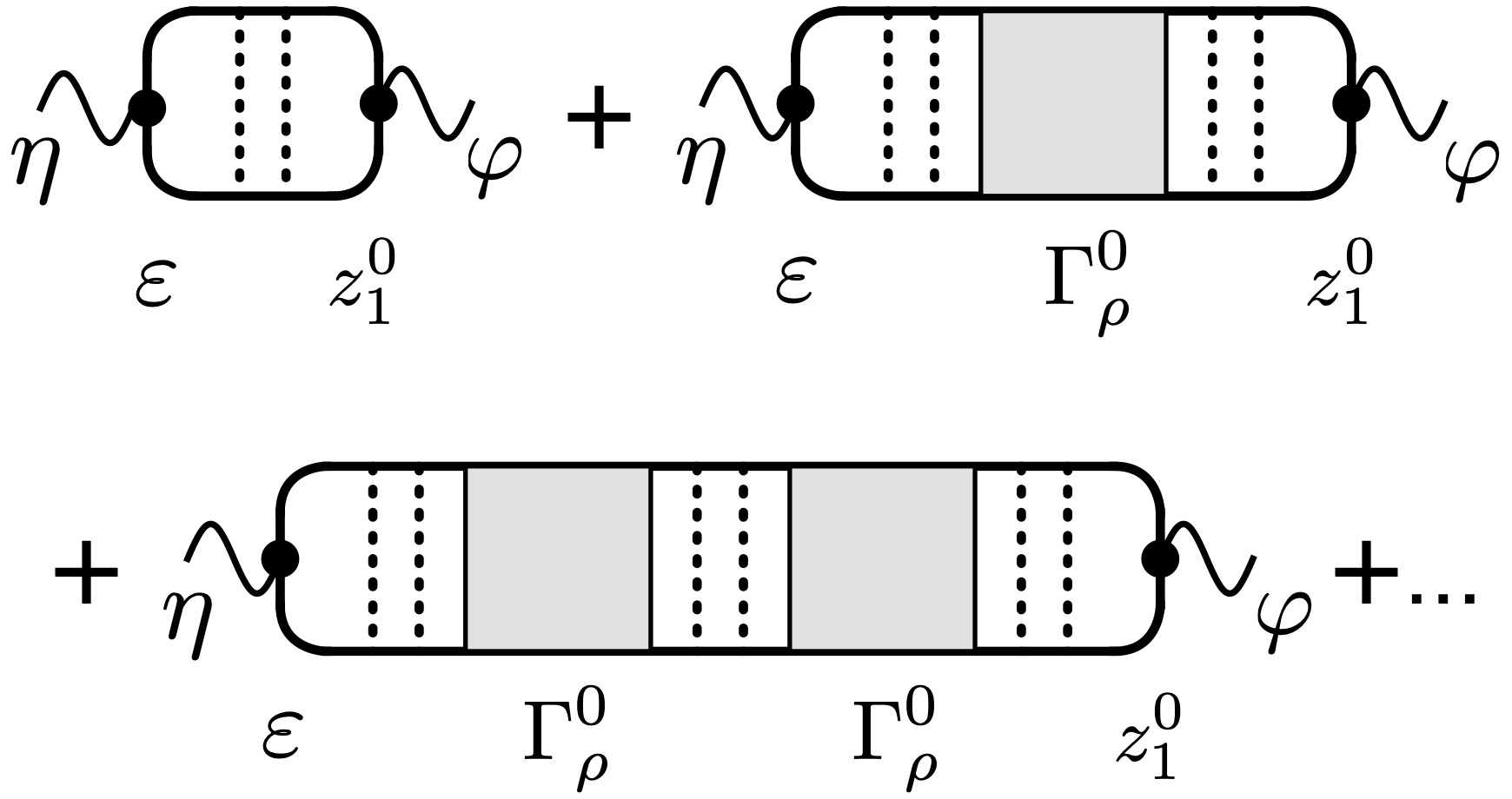}
\caption{Ladder diagrams contributing to $\chi_{kn,1}^{dyn,0}(\bfq,\omega)$. The interaction amplitude $\Gamma_\rho^0$ is related to the Fermi-liquid parameter $F_0^\rho$ as $\Gamma_\rho^0=F_0^\rho/(1+F_0^\rho)$. The scalar potential $\varphi$ and gravitational potential $\eta$, formally introduced in Sec.~\ref{sec:InteractingNLSM}, are source fields coupling to the density and heat density, respectively. The ladders of dotted (impurity) lines represent diffusons.}
\label{fig:heatchargedensity}
\end{figure}

\begin{figure}[tb]
\includegraphics[width=8.2cm]{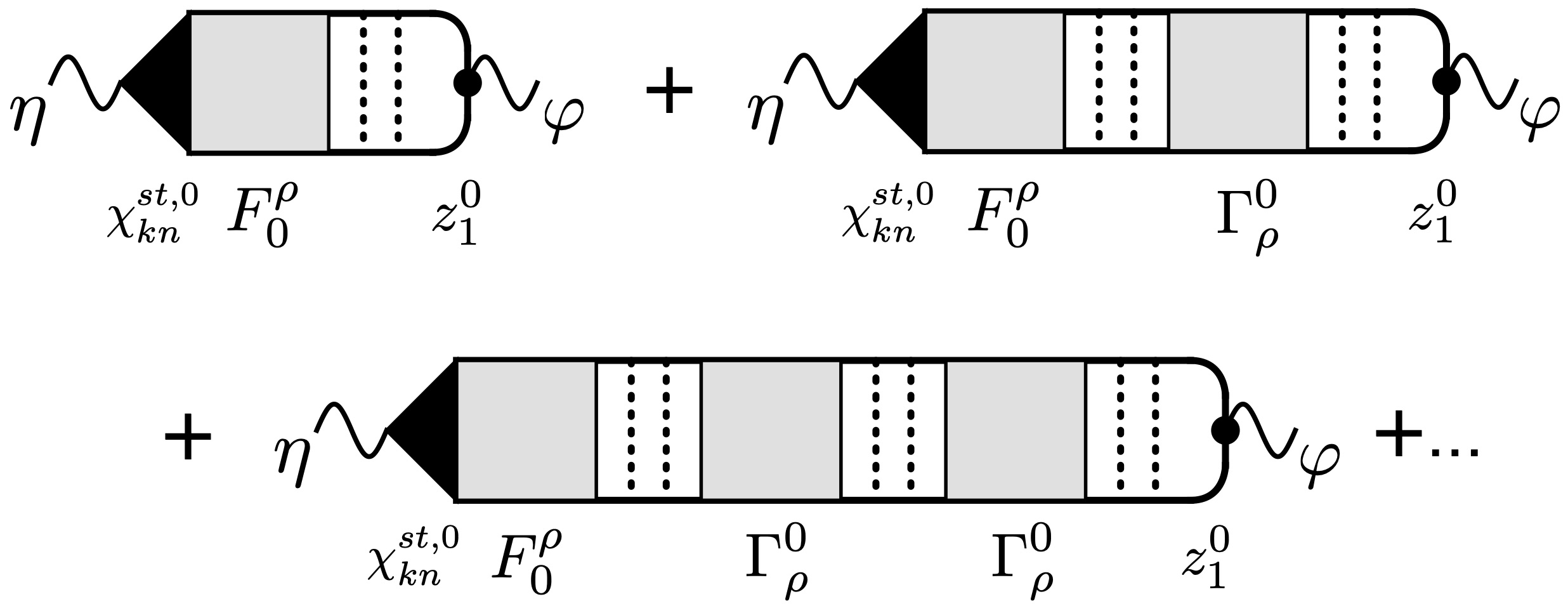}
\caption{Ladder diagrams contributing to $\chi_{kn,2}^{dyn,0}(\bfq,\omega)$.}
\label{fig:etaX}
\end{figure}

The two diagrams relevant for the calculation of the dynamical part of the correlation function are displayed in Fig.~\ref{fig:heatchargedensity} and Fig.~\ref{fig:etaX}. The diagram in Fig.~\ref{fig:heatchargedensity} has two external vertices, 
one vertex symbolizing the heat density coupling to the gravitational potential $\eta$, and a second one for the density coupling to the scalar potential $\varphi$. In general, the heat density is represented by two distinct types of vertices in a diagrammatic representation, one associated with the of the electrons, and the other one associated with the interaction \cite{Schwiete14b}. Postponing a more detailed discussion to Sec.~\ref{sec:InteractingNLSM}, the frequency vertex corresponds to the term $\bar{\psi}(p_+)\eps\psi(p_-)$ in the action, where $p_\pm=(\bfp\pm\bfq/2,\eps\pm\omega/2)$. For the calculation of the dynamical part in the ladder approximation, as in Fig.~\ref{fig:heatchargedensity}, only the frequency vertex is relevant, since the interaction vertex is automatically associated with a loop integration over the frequencies and momenta of the diffusion modes. The analytical expression corresponding to Fig.~\ref{fig:heatchargedensity} reads as
\begin{align}
\chi^{dyn,0}_{kn,1}(\bfq,\omega)&=-2i\pi z_1^0 \frac{\mathcal{D}_1(\bfq,\omega)}{\mathcal{D}(\bfq,\omega)}\int_\eps  \eps\Delta_{\eps_1\eps_2}\bar{\nu}_{\eps}\mathcal{D}_\eps(\bfq,\omega).\label{eq:chiprime}
\end{align}
This equation contains the diffusons $\mathcal{D}(\bfq,\omega)=(D\bfq^2-i\omega)^{-1}$ and $\mathcal{D}_1(\bfq,\omega)=(D\bfq^2-iz_1^0\omega)^{-1}$, where $D$ is the diffusion coefficient. These diffusons appear frequently in the theory of disordered interacting systems \cite{Finkelstein83,Finkelstein90}. For our proposes, it is necessary to introduce another type of diffuson with a frequency dependent diffusion coefficient,
\begin{align}
\mathcal{D}_\eps(\bfq,\omega)=\frac{1}{D_\eps\bfq^2-i\omega},
\end{align}
where $D_\eps=D+\delta D_\eps$. The dominant $\eps$ dependence of $\delta D_\eps$ in two dimensions is given by $\delta D_\eps={\tau \eps}/{m}$, where $\tau$ is the scattering time and $m$ the electron mass. The frequency integral in Eq.~\eqref{eq:chiprime} originates from the product of retarded and advanced Green's functions adjacent to the frequency vertex. The window function $\Delta_{\eps_1\eps_2}=\mathcal{F}_{\eps_1}-\mathcal{F}_{\eps_2}$, where $\mathcal{F}=\tanh(\eps/2T)$ and $\eps_{1/2}=\eps\pm\omega/2$, restricts the $\eps$ integration to a window of order $\omega$ at low temperatures. In the absence of particle-hole asymmetry, i.e., for a constant density and constant diffusion coefficient, the $\eps$ integral in Eq.~\eqref{eq:chiprime} vanishes, because the integrand is odd in $\eps$. After expanding the product $\overline{\nu}_\eps \mathcal{D}_\eps$ to first order in $\eps$, we find that $\chi_{kn,1}^{dyn,0}$ can be presented in the form 
\begin{align}
\chi^{dyn,0}_{kn,1}(\bfq,\omega)&=-T\left(\frac{c_{0,\eps}i\omega}{D_\eps\bfq^2-i\omega}\right)'\frac{D\bfq^2-i\omega}{D_{FL}\bfq^2-i\omega},
\end{align}
where $D_{FL}=D/z_1^0$ and we used the relation $\int_\eps \eps^2\Delta_{\eps_1,\eps_2}=\pi T^2\omega/3$. 

The second contribution to the dynamical part of the correlation function, $\chi^{dyn,0}_{kn,2}$, can be written as 
\begin{align}
\chi_{kn,2}^{dyn,0}(\bfq,\omega)&=\frac{\chi^{st,0}_{kn}F_0^\rho\chi_{nn}^{dyn,0}(\bfq,\omega)}{2\nu z_1^0},\label{eq:chiknetaX}
\end{align}
where $\chi^{dyn,0}_{nn}$ is the dynamical part of the density-density correlation function in the ladder approximation. The structure of the density-density correlation function is well known. In the ladder approximation, its dynamical part is given by $\chi^{dyn,0}_{nn}=-2\nu z_1^0 i\omega/(D_{FL}\bfq^2-i\omega)$, where $\nu$ is the single particle density of states. Figure~\ref{fig:etaX} shows a diagrammatic representation of ${\chi}^{dyn,0}_{kn,2}$. In this diagram, the heat density vertex is connected to the interaction amplitude $F_0^\rho$ by a product of two retarded or two advanced Green's functions. This block, which is depicted as a black triangle, is the origin of the factor $\chi^{st,0}_{kn}$ in Eq.~\eqref{eq:chiknetaX}. The contribution  to the dynamical part of the correlation function $\chi_{kn,2}^{dyn,0}(\bfq,\omega)$ can be combined with the static part to give 
\begin{align}
\chi^{st,0}_{kn}+\chi^{dyn,0}_{kn,2}(\bfq,\omega)=-Tc_{0,\eps}'\frac{D\bfq^2-i\omega}{D_{FL}\bfq^2-i\omega}
\end{align}

The final result for the correlation function in the ladder approximation, which accounts for the non-constant density of states and non-constant diffusion coefficient, and comprises the two contributions to the dynamical part displayed in Eqs.~\eqref{eq:chiprime} and \eqref{eq:chiknetaX}, as well as the static part shown in Eq.~\eqref{eq:chst}, reads as
\begin{align}
\chi^{0}_{kn}&=-T\left(\frac{c_{0,\eps} D_\eps \bfq^2}{D_\eps\bfq^2-i\omega}\right)'\frac{D\bfq^2-i\omega}{D_{FL}\bfq^2-i\omega}.\label{eq:chidynnonconstant}
\end{align}

We see that Eq.~\eqref{eq:chidynnonconstant} is consistent with the general form of $\chi_{kn}$ introduced in Eq.~\eqref{eq:chikngeneral}, after identifying $D_{n}=D_{FL}$ and $D_k=D$, and also with Ref.~\cite{Fabrizio91}. The relations for $D_{n}$ and $D_k$ are also consistent with the known results for the density-density and heat density-heat density correlation functions. By comparison with Eq.~\eqref{eq:chikngeneral}, one obtains the thermoelectric transport coefficient as $L=T(c_{0,\eps} D_\eps)'$, in agreement with the Boltzmann result \cite{Ziman01}.

\subsection{The role of $\bar{\nu}_\eps'$ for the calculation of $L$}

The thermoelectric transport coefficient obtained in the previous section can be written as the sum of two terms, $L=Tc_0 D(\bar{\nu}_\eps'/\nu+D_\eps'/D)$. The two potential sources of particle-hole asymmetry in our model are the $\eps$-dependences of $\bar{\nu}_\eps$ and $D_\eps$. We see that unlike the electric and thermal conductivities, the thermoelectric transport coefficient vanishes when particle-hole asymmetry is neglected. An important observation is that in two dimensions $D_\eps'/D$ is larger by a factor $\eps_F\tau$ compared to $\bar{\nu}_\eps'/\nu$ (for details, see Sec.~\ref{sec:nonconstant}). This is why we may take the density of states as constant when calculating the dominant contribution to the thermoelectric transport coefficient. We demonstrated this explicitly for the ladder approximation. The argument also carries over to the calculation of interaction corrections. The calculation of interaction corrections is typically organized according to the number of loop integrations over slow momenta. Each loop integration generates an additional power of the dimensionless resistance, which serves as a small parameter in the theory. At each given order, corrections proportional to $D_\eps'$ are larger than those proportional to $\bar{\nu}'_\eps$. The main outcome of this discussion is that in two dimensions the leading contributions to the thermoelectric transport coefficient can be calculated by neglecting the $\eps$-dependence of $\bar{\nu}_\eps$. This is the reason why we will restrict ourselves to the constant density of states approximation when deriving the NL$\sigma$M for the calculation of $L$ below.

Before proceeding with the NL$\sigma$M approach, we would like to comment on a subtle point concerning the diffusion coefficient. Besides the $\eps$-dependence originating from $\delta D_\eps =\eps\tau/m$, there is an additional dependence originating from the $\eps$-dependence of the scattering rate $\tau$. Indeed, in the model under consideration a non-constant density of states $\overline{\nu}_\eps$ also goes hand in hand with a non-constant scattering rate $\tau_\eps$. As we will discuss in Sec.~\ref{sec:nonconstant}, the self-consistent Born approximation for the disorder induced self-energy results in the relation $\tau_{\eps}\bar{\nu}_\eps=\mbox{const.}$. This is why the diffusion coefficient acquires an additional frequency dependence through the scattering time. However, this dependence is much weaker than the one originating from the explicit $\eps$-dependence of $\delta D_\eps$.

\section{The generalized NL$\sigma$M}
\label{sec:genNLSM}

The calculation of quantities that are strongly affected by particle-hole asymmetry requires a generalization of the conventional NL$\sigma$M formalism. Indeed, in the Finkel'stein model the density of states and the diffusion coefficient are frequency-independent constants. We will now introduce a generalized NL$\sigma$M that incorporates a frequency-dependent diffusion coefficient. Motivated by the analysis of the heat density-density correlation function in Sec.~\ref{sec:hdcf}, according to which the frequency-dependence of the density of states results in sub-leading corrections to the thermoelectric transport coefficient in two dimensions, the density of states in the generalized model is treated as a constant. The NL$\sigma$M introduced below also contains gravitational and scalar potentials in order to prepare the calculation of the heat density-density correlation function. The derivation of the generalized model will be presented separately in Sec.~\ref{sec:NLSMderivation}. As a first consistency check, we will use the model to reproduce the result for $\chi^0_{kn}$ stated in Eq.~\eqref{eq:chidynnonconstant} for the case of a constant density of states.

\subsection{The NL$\sigma$M action}
\label{subsec:NLSMaction}

The action of the generalized NL$\sigma$M can be written as 
\begin{align}
S^0_{\delta Q}&=S_{0,\eta \varphi}+S_{int,\eta}+S_{\eta}+S_{\varphi}+S_{M}.\label{eq:NLSMgeneral}
\end{align}
The first term on the right hand side is a generalization of the Keldysh NL$\sigma$M for non-interacting systems, and reads as
\begin{align}
S_{0,\eta\varphi}=&\frac{i \pi\nu }{4}\Tr\left[ D(\nabla \underline{\hat{X}})^2+4i \hat{\eps}^\eta_{\varphi_{FL}} \underline{\delta \hat{X}} \right]. \label{eq:S0etaphiX}
\end{align}
Here, $\hat{X}$ is related to the $\hat{Q}$-field used in the conventional sigma model approach as
\begin{align}
\hat{X}=\hat{Q}+\frac{1}{4i}D_\eps'(\nabla{\hat{Q}})^2 \label{eq:X}.
\end{align}
The field $\hat{Q}(\bf r)$ is a matrix in Keldysh space, which also carries two spin and two frequency indices, and fulfills the constraint $\hat{Q}^2=1$. It is understood that the form of the action $S_{0,\eta\varphi}$ displayed in Eq.~\eqref{eq:S0etaphiX} is accurate up to linear order in $D'_\eps$ only. The trace operation $\Tr$ in Eq.~\eqref{eq:S0etaphiX} accounts for all these degrees of freedom and also includes an integration over the coordinates ${\bf r}$. The generalized frequency operator $\hat{\eps}_{\varphi_{FL}}^\eta$ includes source fields and is defined as
\begin{align}
\hat{\eps}^\eta_{\varphi_{FL}}&=\frac{1}{2}\{\hat{\eps}-\hat{\varphi}_{FL},\hat{\lambda}\}.
\end{align}
In this relation, the Fermi liquid renormalization of the density vertex is encoded in $\hat{\varphi}_{FL}=\hat{\varphi}/(1+F_0^\rho)$, and $\hat{\lambda}=(1+\hat{\eta})^{-1}$ contains the gravitational potential $\hat{\eta}$. The frequency operator $\hat{\eps}$ acts as $(\hat{\eps}\hat{Q})_{\eps_1\eps_2}=\eps_1\hat{Q}_{\eps_1\eps_2}$. The matrix structure of the source fields in Keldysh space is defined as follows: $\hat{\varphi}^l=\Sigma_{k=1,2}\varphi_k^l\hat{\gamma}_k$ (and, correspondingly, for $\hat{\lambda}$ and $\hat{\eta}$), where $\hat{\gamma}_1=\hat{\sigma}_0$ and $\hat{\gamma}_2=\hat{\sigma}_1$ are Pauli matrices in Keldysh space. In addition, $\hat{\varphi}$ is also a matrix in frequency space, according to $(\hat{\varphi}_{\bfr})_{\eps\eps'}=\hat{\varphi}_{\bfr,\eps-\eps'}$, and the same applies to $\hat{\lambda}$ and $\hat{\eta}$. The matrix $\hat{Q}$ takes the form $\hat{Q}= \hat{U}\hat{\sigma}_3\hat{\bar{U}}$, where $\hat{U}\hat{\bar{U}}=1$ and $\hat{\sigma}_3$ is the third Pauli matrix in Keldysh space. The second term in Eq.~\eqref{eq:S0etaphiX} contains $\delta \hat{X}=\hat{X}-\hat{\sigma}_3$ in the form of $\underline{\delta \hat{X}}=\hat{u}\delta \hat{X}\hat{u}$. Here, the matrix $\hat{u}$ contains information about the occupation of states,
\begin{align}
\hat{u}_\eps=\left(\ba{cc} 1&\mathcal{F}_\eps\\ 0&-1\ea\right), \quad \hat{u}_\eps=\hat{u}_\eps^{-1}.\label{eq:defu}
\end{align}

In order to make contact with the conventional model, it is useful to write the action $S_{0,\eta,\varphi}$ displayed in Eq.~\eqref{eq:S0etaphiX} in terms of the matrix field $\hat{Q}$ as $S_{0,\eta,\varphi}=S_{0,\eta,\varphi}^{(1)}+S_{0,\eta,\varphi}^{(2)}$, where
\begin{align}
S^{(1)}_{0,\eta\varphi}=&\frac{i \pi\nu }{4}\Tr\left[ \hat{D}_{\hat{\eps}^\eta_{\varphi_{FL}}}(\nabla \underline{\hat{Q}})^2+4i \hat{\eps}^\eta_{\varphi_{FL}} \underline{\delta \hat{Q}} \right],\label{eq:S0etaphi}\\
S^{(2)}_{0,\eta\varphi}=&-\frac{\pi\nu}{8}DD_\eps'\Tr[\nabla^2 \hat{Q}(\nabla \hat{Q})^2].\no
\end{align}
It is instructive to first discuss the form of $S_{0,\eta\varphi}$ in the absence of the source fields $\varphi$ and $\eta$ (for $\hat{\eps}^\eta_{\varphi_{FL}}\rightarrow \hat{\eps}$). Then, the action differs from the conventional Keldysh NL$\sigma$M for non-interacting system by (i) the $\hat{\eps}$ dependence of the diffusion coefficient, $\hat{D}_{\hat{\eps}}=D+\frac{\tau}{m} \hat{\eps}$, and (ii) the presence of the higher order gradient term. We already saw in Sec.~\ref{sec:hdcf} that the frequency dependence of the diffusion coefficient is crucial for the calculation of the thermoelectric transport coefficient.  The term $S_M$ in Eq.~\eqref{eq:NLSMgeneral} has the same structure as $S^{(2)}_{0,\eta\varphi}$, but a different numerical coefficient, 
\begin{align}
S_M=\frac{\pi\nu}{16}DD_\eps'\Tr[\nabla^2 \hat{Q}(\nabla \hat{Q})^2].\label{eq:SMmain}
\end{align}
Despite the structural similarity, the two terms $S^{(2)}_{0,\eta\varphi}$ and $S_M$ have a rather different origin. While $S^{(2)}_{0,\eta\varphi}$ is derived by exclusively focusing on the soft modes, $S_M$ is obtained via an explicit integration over massive modes (see Appendix \ref{app:SM} and Ref. \cite{Wang94}). By combining $S^{(2)}_{0,\eta\varphi}$ and $S_M$, one obtains the total four-gradient term in the action as 
\begin{align}
S^{(2)}_{0,\eta\varphi}+S_M=-\frac{\pi\nu}{16}DD_\eps'\Tr[\nabla^2 \hat{Q}(\nabla \hat{Q})^2].
\end{align} 

The importance of the higher gradient term in the action does not reveal itself in the ladder approximation for the heat density-density correlation function. This term gives rise to (generalized) Hikami-box diagrams, however, and is therefore expected to become important for the calculation of interaction corrections to the dynamical part of the correlation function.

The interaction term in the action takes the form
\begin{align}
S_{int,\eta}=-\frac{\pi^2\nu}{8}\int_{\bfr,\eps_i} \Big(\tr[\hat{\gamma}_i(\hat{\lambda}\underline{\delta \hat{X}})_{\eps_1\eps_2}]\hat{\gamma}_2^{ij}\Gamma_\rho^0 \tr[\hat{\gamma}_j \underline{\delta \hat{X}}_{\eps_3\eps_4}]\no\\
+\tr[\hat{\gamma}_i\bfsigma (\hat{\lambda}\underline{\delta \hat{X}})_{\eps_1\eps_2}]\hat{\gamma}_2^{ij}\Gamma_\sigma^0\tr[\hat{\gamma}_j\bfsigma \underline{\delta \hat{X}}_{\eps_3\eps_4}]\Big)\delta_{\eps_1-\eps_2,\eps_4-\eps_3},\label{eq:slambda}
\end{align}
where we abbreviated $\delta_{\eps,\eps'}=2\pi\delta(\eps-\eps')$. The amplitudes $\Gamma_{\rho/\sigma}^0$ are related to the Fermi liquid amplitudes as follows: $\Gamma_{\rho}^0=F_0^\rho/(1+F_0^\rho)$ and $\Gamma_{\sigma}^0=F_0^\sigma/(1+F_0^\sigma)$. The presence of $\hat{\lambda}$ in the action \eqref{eq:slambda} reflects the fact that the heat density includes a contribution from the interaction itself. An analogous term was found in the context of the NL$\sigma$M approach to the calculation of the heat density-heat density correlation function in Refs.~\cite{Schwiete14a,Schwiete14b}, albeit with the matrix $\hat{Q}$ instead of $\hat{X}$.

As far as the last two contributions to $S_{\delta Q}^0$ in Eq.~\eqref{eq:NLSMgeneral} are concerned, they describe the coupling of the quantum component of the potentials, $\varphi_2$ and $\eta_2$, to the non-interacting density $n_0$ and heat density $k_0$, respectively 
\begin{align}
S_\eta&=-2k_0\int_x\eta_2(x),\quad S_\varphi=-2n_0\int_x\varphi_2(x).
\end{align}

The gravitational potentials $\eta_1$ and $\eta_2$ \cite{Luttinger64,Shastry09} in combination with the scalar potentials $\varphi_1$ and $\varphi_2$ allow us to formulate a linear response theory for the thermoelectric transport. The correlation functions $\chi_{kn}$ in the diffusive limit can be obtained from the Keldysh partition function $\mathcal{Z}=\int DQ \exp(iS_{\delta Q}^0)$ as
\be
\chi_{kn}(x_1,x_2)&=&\left.\frac{i}{2}\frac{\delta^2 \mathcal{Z}}{\delta \eta_2(x_1)\delta \varphi_1(x_2)}\right|_{\vec{\eta}=\vec{\varphi}=0},\label{eq:chisources}
\ee
while $\chi_{nk}$ can be found by switching the roles of $\eta$ and $\varphi$. One can also calculate the heat density as  
\begin{align}
\langle \hat{k}\rangle_T&=\left\langle k_{cl}(x)\right\rangle=\frac{i}{2}\left.\frac{\delta{\mathcal{Z}}}{\delta \eta_2(x)}\right|_{\vec{\eta}=\vec{\varphi}=0}\label{eq:kav},
\end{align}
and the charge density by differentiating with respect to $\varphi_2$. In these equations, we wrote $\vec{\eta}=(\eta_1,\eta_2)^T$ and $\vec{\varphi}=(\varphi_1,\varphi_2)^T$. 

Equations~\eqref{eq:S0etaphiX},~\eqref{eq:SMmain} and \eqref{eq:slambda} are the main results of this manuscript. We see that particle-hole asymmetry can be incorporated into the conventional NL$\sigma$M through the replacement $\hat{Q}\rightarrow \hat{X}$, and the addition of the term $S_M$. It is worth noting that the effective potential $\hat{\varphi}_{FL}$ couples to the field $\delta \hat{X}$ in the action. Therefore, $\delta \hat{X}$ describes density fluctuations.

\subsection{Particle-hole asymmetry}
\label{sec:pha}

In the absence of sources and for a constant diffusion coefficient, Eq.~\eqref{eq:NLSMgeneral} reduces to the conventional Keldysh NL$\sigma$M action $S_F[\hat{Q}]$ for interacting systems in the unitary symmetry class. This action is invariant under a certain transformation of the $\hat{Q}$ matrices. Indeed, $S_F[\hat{Q}]=S_F[\hat{Q}']$ holds for
\begin{align}
\hat{Q}'_{\eps_1\eps_2}=-\sigma_2\hat{\sigma}_1\hat{Q}^t_{-\eps_1,-\eps_2}\hat{\sigma}_1\sigma_2,\label{eq:trafo}
\end{align}
where $\hat{\sigma}_1$ is a Pauli matrix in Keldysh space, $\sigma_2$ acts in spin space and the transposition in $\hat{Q}^t$ operates on Keldysh space, on spin indices, and on frequencies [a similar transformation was used in Ref.~\cite{Wang94}]. The saddle point matrix $\hat{\sigma}_3$ and the matrix $\hat{u}$ are also invariant under this transformation, $\hat{\sigma}_3'=\hat{\sigma}_3$, $\hat{u}'=\hat{u}$. 

To understand this observation it is instructive to study the effect of the transformation \eqref{eq:trafo} on the electronic Green's function, which is connected to $\hat{Q}$ via the saddle point equation. When applying this transformation to $\hat{G}_0=\mbox{diag}(G_0^R,G_0^A)$, where $G_0^R$ and $G_0^A$ are retarded and advanced Green's functions and to the equilibrium Keldysh Green's function $\underline{\hat{G}}=\hat{u}\hat{G}_0\hat{u}$, the ``particle" Hamiltonian $\hat{k}=\hat{h}_0-\mu$ transforms into the ``hole" Hamiltonian $\hat{k}'=-(i\sigma_2)\hat{k}^T(i\sigma_2)^{-1}$ by a combination of a sign change and a time reversal operation \cite{Transposition}. The invariance of the conventional NL$\sigma$M action under the particle-hole transformation \eqref{eq:trafo} is a consequence of the approximations used during the derivation, for which the velocity and the density of states are treated as constant. The approximation of a constant velocity has been avoided for the derivation of the generalized model, and this is why \eqref{eq:NLSMgeneral} incorporates particle-hole asymmetry.

The generalized NL$\sigma$M action can be broken down to individual pieces that are either even, $S_i[\hat{Q}']=S_i[\hat{Q}]$, or odd, $S_i[\hat{Q}']=-S_i[\hat{Q}]$, under the transformation \eqref{eq:trafo}. As already noted, the conventional NL$\sigma$M action $S_F$, which is obtained from $S^0_{\delta Q}$ by setting $\eta=\varphi=D_\eps'=0$, is even under transformation \eqref{eq:trafo}. The source terms for $D_\eps'=0$ transform as  
\begin{align}
\Tr[\{\hat{\eps}-\hat{\varphi}_{FL},\hat{\lambda}\}\delta \hat{Q}']=\Tr[\{\hat{\eps}+\hat{\varphi}_{FL},\hat{\lambda}\}\delta \hat{Q}],\label{eq:trafosources}
\end{align}
and $S_{int,\eta}[\hat{Q}']=S_{int,\eta}[\hat{Q}]$. Since the generalized NL$\sigma$M is obtained from the conventional one by the replacement $\hat{Q}\rightarrow \hat{X}=\hat{Q}+\frac{1}{4i}D_\eps'(\nabla \hat{Q})^2$ (up to the change in coefficient for the four-gradient term), each term in the conventional model has a partner containing $D_\eps'$. One can see that the terms with $D_\eps'$ acquire an additional minus sign under the transformation $\hat{Q}\rightarrow \hat{Q}'$. 

These observations have important consequences for the calculation of correlation functions. The heat density-heat density and the density-density correlation function are obtained as second derivatives of the Keldysh partition function with respect to the source fields $\eta$ and $\varphi$, respectively. Therefore, the sign change in Eq.~\eqref{eq:trafosources} is not relevant and the inclusion of particle-hole asymmetry is not required for the calculation of these correlation functions. By contrast, the calculation of the heat density-density correlation function $\chi_{kn}$ requires derivatives with respect to both $\varphi$ and $\eta$ [compare Eq.~\eqref{eq:chisources}]. Due to the sign change in the source term in Eq.~\eqref{eq:trafosources}, a non-vanishing result for $\chi_{kn}$ can only be obtained by including terms with $D_\eps'\ne 0$, i.e., with particle-hole asymmetry. 

The symmetry considerations presented above are a valuable guide for the derivation of the generalized NL$\sigma$M. They allow us to distinguish terms that share the symmetry of the conventional NL$\sigma$M from those terms that change the symmetry and therefore need to be included into the generalized model.

\subsection{$\chi_{kn}^0$ from the generalized NL$\sigma$M}

We will now discuss how the correlation function $\chi_{kn}$ in the ladder approximation can be obtained from the generalized NL$\sigma$M. In the absence of interaction corrections, the static part of the correlation function vanishes and only the dynamical part needs to be considered. This is an immediate consequence of the constant-density approximation used for the derivation of the NL$\sigma$M. Interaction corrections arising from momentum and frequency integrations over diffusion modes are neglected in the ladder approximation. The ladder can be obtained by treating fluctuations near the saddle point $\hat{Q}=\hat{\sigma}_3$ in the NL$\sigma$M in the Gaussian approximation. In order to derive the Gaussian action, one needs to choose a parameterization for the matrix $\hat{U}$. In this paper, we will work with the exponential parameterization, for which $\hat{U}=\mbox{e}^{-\hat{P}/2}$ with the additional constraint $\{\hat{P},\hat{\sigma}_3\}=0$. Details concerning the parametrization, and the contraction rules for averages with respect to the Gaussian action are presented in Appendix \ref{app:perturbations}. 

The calculation is simplified by the fact that vertices originating from the interaction part $S_{int,\eta}$ or the gradient term in $S_{0,\eta\varphi}$ do not contribute in the ladder approximation. Only the source fields contained in the second term of the action $S_{0,\eta\varphi}$ in Eq.~\eqref{eq:S0etaphi} and a first order expansion of $\delta \hat{Q}$ in $\hat{P}$ are relevant for the vertices. Using the contraction rule \eqref{eq:dsinglet} stated in appendix~\ref{app:perturbations}, one obtains
\begin{align}
\tilde{\chi}_{kn,1}^{dyn,0}(\bfq,\omega)&=-2i\pi\nu z_1^0 \frac{\mathcal{D}_1(\bfq,\omega)}{\mathcal{D}(\bfq,\omega)}\int_\eps \eps\Delta_{12}\mathcal{D}_\eps(\bfq,\omega).\label{eq:tildechikn1}
\end{align}
The diagrammatic representation is shown in Fig.~\ref{fig:heatchargedensity}. By comparison with Eq.~\eqref{eq:chiprime}, we see immediately that $\tilde{\chi}_{kn,1}^{dyn,0}$ is obtained from $\chi_{kn,1}^{dyn,0}$ in the constant density of states approximation, as expected. The contribution $\chi_{kn,2}^{dyn,0}$ of Eq.~\eqref{eq:chiknetaX} has no analog here because it is proportional to $\overline{\nu}_{\eps}'$. [In Sec.~\ref{sec:nonconstant} below, we will explain how this term can included into the generalized NL$\sigma$M.]  Since the static part of the correlation function also vanishes in the constant density of states approximation, Eq.~\eqref{eq:tildechikn1} represents the only non-vanishing contribution to $\tilde{\chi}^0_{kn}$. After performing the integration, the result can be written in the form
\begin{align}
\tilde{\chi}^{0}_{kn}(\bfq,\omega)
&=-T c_0\partial_\eps\left[\frac{D_\eps \bfq^2}{D_\eps\bfq^2-i\omega}\right]\frac{D\bfq^2-i\omega}{D_{FL}\bfq^2 -i\omega}.\label{eq:chidynconst}
\end{align}
This result can also be obtained from the more general Eq.~\eqref{eq:chidynnonconstant} in the limit $c_{0,\eps}\rightarrow c_0=2\pi^2 \nu T/3$, i.e. for a constant density of states. By comparing the result $\tilde{\chi}_{kn}^0(\bfq,\omega)$ to the general form of the correlation function stated in Eq.~\eqref{eq:chikngeneral}, we find $L=Tc_0D'$ [and, obviously, $\chi_{kn}^{st,0}=0$]. The resulting Seebeck coefficient is $eS=\pi^2 T/3\eps_F$, in agreement with the Boltzmann result for two-dimensional systems. Interestingly, in the absence of interaction corrections, the result is independent of the interactions.

\section{Interaction corrections to the static part of the correlation function}
\label{sec:Heatdens}

As a further application of the NL$\sigma$M formalism, we calculate interaction corrections to the static part of the heat density-density correlation function, $\chi_{kn}^{st}$. The structure of the NL$\sigma$M action \eqref{eq:NLSMgeneral} in the presence of source fields is quite intricate. This is why we calculate the interaction corrections to the static part in two different ways, which will provide a valuable test of the structure of the model. First, in Sec.~\ref{subsec:hd}, we make use of the relation of the static part to certain thermodynamic susceptibilities. Starting point for this approach is the relation
\begin{align}
\chi_{kn}^{st}=&\frac{T}{V}\partial_T\partial_\mu\Omega,
\end{align}
where $\Omega$ is the grand canonical potential. The second derivative on the right hand side may be interpreted in two ways, 
\begin{align}
\frac{T}{V}\partial_T\partial_\mu\Omega=&-T\partial_T \langle \hat{n}\rangle_T=-\partial_\mu\langle \hat{k}\rangle_T-\langle \hat{n}\rangle_T.\label{eq:Maxwell}
\end{align}
The thermal averages $\langle \hat{k} \rangle_T$ and $\langle \hat{n}\rangle_T$ can be calculated straightforwardly in our formalism, as explained in Sec.~\ref{subsec:NLSMaction}. As a byproduct, we will verify that the Maxwell relation stated on the right hand side of Eq.~\eqref{eq:Maxwell} is reproduced. This first approach relies on terms of first order in the source fields, see Eqs.~\eqref{eq:kav}. As an alternative route to the calculation of the correlation function, we will study the limit $\chi^{st}_{kn}=\chi_{kn}({\bf q}\rightarrow 0, \omega=0)$ directly in Sec.~\ref{sec:stat}. This approach makes use of terms that are of second order in the source fields, see Eq.~\eqref{eq:chisources}.

\subsection{Calculation of interaction corrections to the thermodynamic susceptibilities}
\label{subsec:hd}

\begin{figure}
\includegraphics[width=8.5cm]{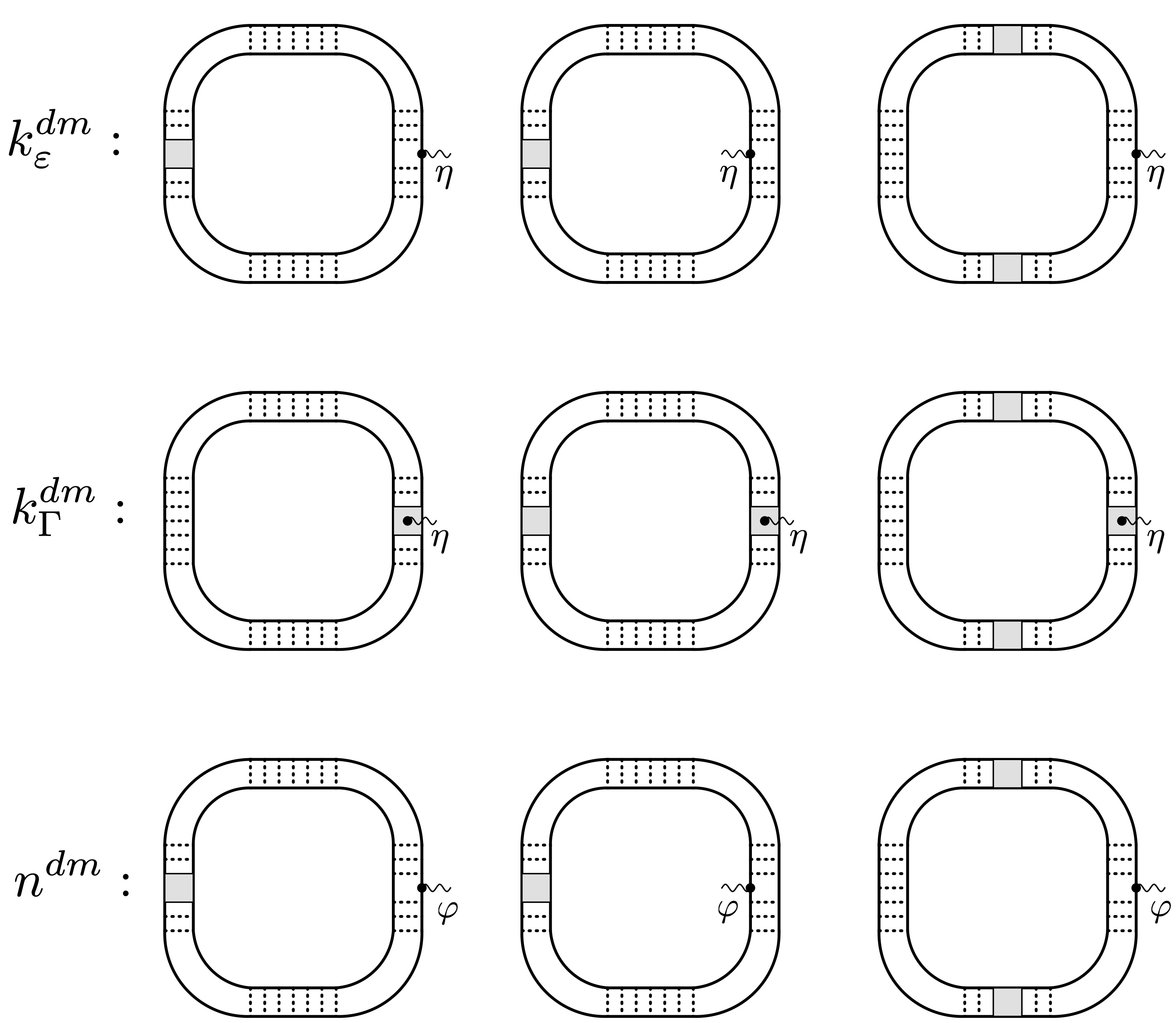}
\caption{The diffusion mode contribution to the heat density and the charge density. Rectangles symbolize the scattering amplitudes; rescattering is either in the singlet channel with amplitudes $\Gamma_\rho$, or in the triplet channel with amplitudes $\Gamma_\sigma$. Ladders of dotted lines stand for the bare diffusons $\mathcal{D}$ or $\mathcal{D}_\eps$. The terms $k^{dm}_{\eps,1}$ and $k^{dm}_{\eps,2}$ are symbolized by the same diagram labeled as $k^{dm}_\eps$.}
\label{fig:dmheatdensity}
\end{figure}

We will first make use of the relation $\chi^{st}_{kn}=-T\partial_T \langle \hat{n}\rangle_T$ for the calculation of the static part of the correlation function. The diffusion mode contribution to the density $n^{dm}$ can be obtained with the help of the NL$\sigma$M action \eqref{eq:NLSMgeneral}. After performing the differentiation with respect to $\varphi_2$, one obtains the expression 
\begin{align}
n^{dm}(x)=-\frac{\pi\nu z^0_1}{2}\tr\left[\hat{\gamma}_2\underline{\delta \hat{X}_{tt}}({\bf r})\right].\label{eq:ndmstart}
\end{align}
Here, and in the following, averaging with respect to the NL$\sigma$M action in the absence of sources is implied. For the perturbative calculation of the diffusion mode contribution, $\delta \hat{X}$ in Eq.~\eqref{eq:kdeltaQ} should be expanded to second order in the generators $\hat{P}$. We split the resulting terms into two parts, $n^{dm}=n^{dm}_1+n^{dm}_2$, where  
\begin{align}
n^{dm}_1&=-\frac{\pi\nu z_1^0}{4}\tr\left[\hat{\gamma}_2\underline{\hat{\sigma}_3\hat{P}^2_{tt}}\right],\\
n^{dm}_2&=-\frac{\pi \nu i z_1^0}{8}D'\tr\left[\hat{\gamma}_2\underline{(\nabla \hat{P})^2_{tt}}\right].
\end{align}
These expressions can be averaged with the help of Eqs.~\eqref{eq:dsinglet} and Eqs.~\eqref{eq:dtriplet}. Fig.~\ref{fig:dmheatdensity} provides a diagrammatic representation of the resulting contributions. The following identities are useful for the calculation,
\be
1-\mathcal{F}_{\eps+\frac{\omega}{2}}\mathcal{F}_{\eps-\frac{\omega}{2}}&=&\mathcal{B}_\omega(\mathcal{F}_{\eps+\frac{\omega}{2}}-\mathcal{F}_{\eps-\frac{\omega}{2}}),\\
\int_\eps (\mathcal{F}_{\eps+\frac{\omega}{2}}-\mathcal{F}_{\eps-\frac{\omega}{2}})&=&\frac{\omega}{\pi},
\ee
and $1-\mathcal{F}^2_\eps=2T\mathcal{F}'_\eps$, where $\mathcal{B_{\omega}}=\coth \frac{\omega}{2T}$ is the bosonic equilibrium distribution function. The calculation shows that only $n^{dm}_2$ gives a contribution, which leads us to 
\begin{align}
n^{dm}&=\frac{z_1^0}{2}\int_{\bfq,\omega} \omega\mathcal{B}_\omega D_\eps'{\bf q}^2\mathcal{D}(\Gamma_\rho^0\mathcal{D}_1+3\Gamma_\sigma^0\mathcal{D}_2).
\label{eq:ndm1}
\end{align}
The relation $\mathcal{D}^{-1}_{1,2}-\mathcal{D}^{-1}=i\omega\Gamma_{\rho,\sigma}$ was used in obtaining this result. Making use of the identity $T\partial_T \mathcal{B}_\omega=-\omega \partial_\omega \mathcal{B}_\omega$, we further obtain
\begin{align}
-T\partial_T n^{dm}&=\frac{z_1^0}{2}\int_{\bfq,\omega} \omega^2\partial_\omega \mathcal{B}_\omega D_\eps'{\bf q}^2\mathcal{D}(\Gamma_\rho^0\mathcal{D}_1+3\Gamma_\sigma^0\mathcal{D}_2).\label{eq:TdTn}
\end{align}
The factor $\partial_\omega \mathcal{B}_\omega$ in Eq.~\eqref{eq:kd} constrains the frequency $\omega$ to be of the order of $T$. This allows us to neglect all frequencies in the diffusion propagators $\mathcal{D}$, $\mathcal{D}_1$ and $\mathcal{D}_2$ in the expression for $-T\partial_T n^{dm}$. The remaining logarithmic in integral in ${\bf q}$ then acquires the coefficient $\Gamma_\rho^0+3\Gamma_\sigma^0$. Using the integral $\int_\omega \omega^2\partial_\omega \mathcal{B}_\omega=-2\pi T^2/3$, and the expression for the specific heat $c_0=2\pi^2 \nu T/3$, we find
\begin{align}
\chi_{kn}^{st,dm}=-T\partial_T n^{dm}=Tz_1^0 c_0\delta z D'/D.
\end{align}
Here, $\delta z$ is the known result for the correction to the frequency renormalization $z$ at first order in the small parameter $\rho\equiv ((2\pi)^2\nu D)^{-1}$, namely $\delta z=-\frac{1}{2} \rho(\Gamma^0_\rho+3\Gamma_\sigma^0) \log{1}/{T\tau}$. Noting that $\partial_\mu \rho=-z_1^0\rho D_\eps'/D$, the result may also be written as $\chi_{kn}^{st,dm}=-Tc_0\partial_\mu z=-T\partial_\mu c$, where $c$ is the specific heat including interaction corrections and the $\mu$ dependence of the density of states has been neglected in the last equality.

We proceed by confirming the Maxwell relation stated in Eq.~\eqref{eq:Maxwell} in the context of the NL$\sigma$M approach. This requires the knowledge of the diffusion mode contribution to the heat density $k^{dm}$. We perform the differentiation of the Keldysh partition function with respect to $\eta_2$ to obtain
\be
&&k^{dm}(x)=
-\frac{\pi\nu i }{4}\tr\left[\hat{\gamma}_2(\partial_{t}-\partial_{t'})_{t'=t}\underline{\delta\hat{X}_{tt'}}({\bf r})\right]\no\\
&&\quad-\frac{\pi^2\nu}{16}\sum_{i=1}^2\sum_{l=0}^3\tr\left[\hat{\gamma}_i\sigma^l\underline{\delta\hat{X}_{tt}}({\bf r})\right]\tr\left[\hat{\gamma}_i\sigma^l\underline{\delta \hat{X}_{tt}}({\bf r})\right]\no\\
&&\quad \times \mbox{diag}\left(\Gamma^0_\rho,\Gamma^0_\sigma,\Gamma^0_\sigma,\Gamma^0_\sigma\right)_{ll}.\label{eq:kdeltaQ}
\ee
As for the calculation of the density, $\delta \hat{X}$ in Eq.~\eqref{eq:kdeltaQ} needs to be expanded to second order in $\hat{P}$. For the sake of the discussion, we distinguish three contributions, 
\be
k^{dm}_{\eps,1}&=&-\frac{\pi\nu i }{8}\tr\left[\hat{\gamma}_2(\partial_{t}-\partial_{t'})_{t'=t}\underline{\hat{\sigma}_3\hat{P}^2_{tt'}}({\bf r})\right],\label{eq:keps1}\\
k^{dm}_{\eps,2}&=&\frac{\pi\nu D' }{16}\tr\left[\hat{\gamma}_2(\partial_{t}-\partial_{t'})_{t'=t}\underline{(\nabla\hat{P})^2_{tt'}}({\bf r})\right],\\
k^{dm}_{\Gamma}&=&-\frac{\pi^2\nu}{16}\sum_{i=1}^2\sum_{k=0}^3\tr\left[\hat{\gamma}_i\sigma_k\underline{\hat{\sigma}_3\hat{P}_{tt}}({\bf r})\right]\tr\left[\hat{\gamma}_i\sigma_k\underline{\hat{\sigma}_3 \hat{P}_{tt}}({\bf r})\right]\no\\
&&\times \mbox{diag}\left(\Gamma^0_\rho,\Gamma^0_\sigma,\Gamma^0_\sigma,\Gamma^0_\sigma\right)_{kk}.\label{eq:kgamma}
\ee
An illustration of these contributions is shown in Fig.~\ref{fig:dmheatdensity}. Upon averaging with the Gaussian action, we obtain the leading contributions as
\be
k^{dm}_{\eps,1}&=&-\frac{1}{2}\int_{\bfq,\omega}\omega\mathcal{B}_\omega\left(\mathcal{D}-\mathcal{D}_1+3(\mathcal{D}-\mathcal{D}_2)\right),\label{eq:kdeps}\\
k^{dm}_\Gamma&=&-\frac{1}{2}\int_{\bfq,\omega} \omega\mathcal{B}_\omega \left(\Gamma^0_\rho\mathcal{D}_1+3\Gamma^0_\sigma\mathcal{D}_2\right),\;\label{eq:kdgamma}
\ee
whereas $k^{dm}_{\eps,2}=0$. After adding these two terms one obtains the total collective mode contribution to the heat density as \cite{Schwiete14b}
\begin{align}
&k^{dm}=k^{dm}_{\eps,1}+k^{dm}_\Gamma\no \\
&=\frac{1}{2}\int_{\bfq,\omega} \omega \mathcal{B}_\omega \left[z_1^0\mathcal{D}_1-\mathcal{D}+3(z_2^0\mathcal{D}_2-\mathcal{D})\right]\label{eq:kd}.
\end{align}
where $\bar{\mathcal{D}}$ is the advanced diffuson. This expression coincides with previously obtained results \cite{Catelani05,Schwiete14b}.

With the results for $n^{dm}$ and $k^{dm}$ at hand, it is convenient to write the Maxwell relation for the diffusion mode contributions in the form 
\begin{align}
(1-T\partial_T)n^{dm}=-\partial_\mu k^{dm}.\label{eq:thermaux}
\end{align}
We first study the left hand side of this equation based on Eq.~\eqref{eq:ndm1}. After using the relation $T\partial_T\mathcal{B}_\omega=-\omega\partial_\omega \mathcal{B}_\omega$, a partial integration in $\omega$ can be performed to find
\begin{align}
&(1-T\partial_T)n^{dm}\label{eq:thermaux1}\\
&=-\frac{z_1^0}{2}\int_{\bfq,\omega} D'\bfq^2\omega\mathcal{B}_\omega (1+\omega\partial_\omega)\left[\mathcal{D}(\Gamma_\rho^0\mathcal{D}_1+3\Gamma_\sigma^0\mathcal{D}_2)\right].\no
\end{align}
We next turn to the right hand side of Eq.~\eqref{eq:thermaux}. The differentiation of $k^{dm}$, Eq.~\eqref{eq:kd}, with respect to $\mu$ can easily be performed with the help of the relation $\partial_\mu D=z_1^0\partial_\eps D_\eps$ \cite{Fabrizio91}. By comparing the result to \eqref{eq:thermaux1}, one establishes the relation \eqref{eq:thermaux}.

Let us briefly comment on a technical aspect of the calculation. For the sake of convenience, physical quantities in this manuscript are expressed through derivatives with respect to the quantum component of the source fields. Eq.~\eqref{eq:kav} is a typical example. In this way, the Keldysh component of the Green's function is generated and, as a consequence, the distribution functions $\mathcal{F}$ and $\mathcal{B}$ enter the integrals \cite{Kamenev11}. Strictly speaking, the lesser component of the Green's function $G^<=(G^K-G^R+G^A)/2$ and, correspondingly, the Fermi distribution $n_F=(1-\mathcal{F})/2$ and Planck distribution $n_P=(\mathcal{B}-1)/2$, should be used instead. When calculating thermodynamic quantities it is sometimes useful to remember the distinction when interpreting the results. For example, upon symmetrization in $\omega$ and replacing $\mathcal{B}\rightarrow 2 n_P$, Eq.~\eqref{eq:kd} reads as 
\begin{align}\
k^{dm}=&\int_{\bfq,\omega>0} \omega n_P(\omega) D\bfq^2\times\\
&\times[z_1\mathcal{D}_1\bar{\mathcal{D}}_1-\mathcal{D}\bar{\mathcal{D}}+3(z_2^0\mathcal{D}_2\bar{\mathcal{D}}_2-\mathcal{D}\bar{\mathcal{D}})],\nonumber
\end{align}
and Eq.~\eqref{eq:thermaux1} should be understood in a similar way.

\subsection{Calculation of interaction corrections to the static part of the correlation function}
\label{sec:stat}

We now present a direct calculation of the static part of the correlation function $\chi^{st}_{kn}$. According to Eq.~\eqref{eq:chisources}, $\chi^{st}_{kn}$ can be obtained by differentiating the heat density $k_{\varphi_1}(x)=(i/2)\delta \mathcal{Z}/\delta \varphi_2(x)$ with respect to the classical component of the scalar potential $\varphi_1$: $\chi_{kn}^{st}=\partial k^{dm}_{\varphi_1}/\partial \varphi_1$, where $\varphi_1$ may be taken as constant. Therefore, $k_{\varphi_1}(x)$ needs to be found in the presence of $\varphi_1$. One obtains $k^{dm}_{\varphi_1}=k^{dm}_{\varphi_1,\eps}+k^{dm}_{\varphi_1,\Gamma}+k^{dm}_{\varphi_1,n}$, where
\begin{align}
k^{dm}_{\varphi_1,\eps}(x)=&
-\frac{\pi\nu i }{4}\tr\left[\hat{\gamma}_2(\partial_{t}-\partial_{t'})_{t'=t}\underline{\delta\hat{X}_{tt'}}({\bf r})\right],\label{eq:kdmphieps}\\
k^{dm}_{\varphi_1,\Gamma}(x)=&-\frac{\pi^2\nu}{16}\sum_{i=1}^2\sum_{l=0}^3\tr\left[\hat{\gamma}_i\sigma^l\underline{\delta\hat{X}_{tt}}({\bf r})\right]\tr\left[\hat{\gamma}_i\sigma^l\underline{\delta \hat{X}_{tt}}({\bf r})\right]\no\\
&\times \mbox{diag}\left(\Gamma_\rho,\Gamma_\sigma,\Gamma_\sigma,\Gamma_\sigma\right)_{ll},\label{eq:kdmphiGamma}\\
k^{dm}_{\varphi_1,n}(x)=&\frac{\pi \nu z_1^0}{2} \tr[\varphi_1\hat{\gamma}_2\underline{\delta \hat{X}_{tt}}({\bf r})].
\end{align}
\begin{figure}[tb]
\includegraphics[width=8.5cm]{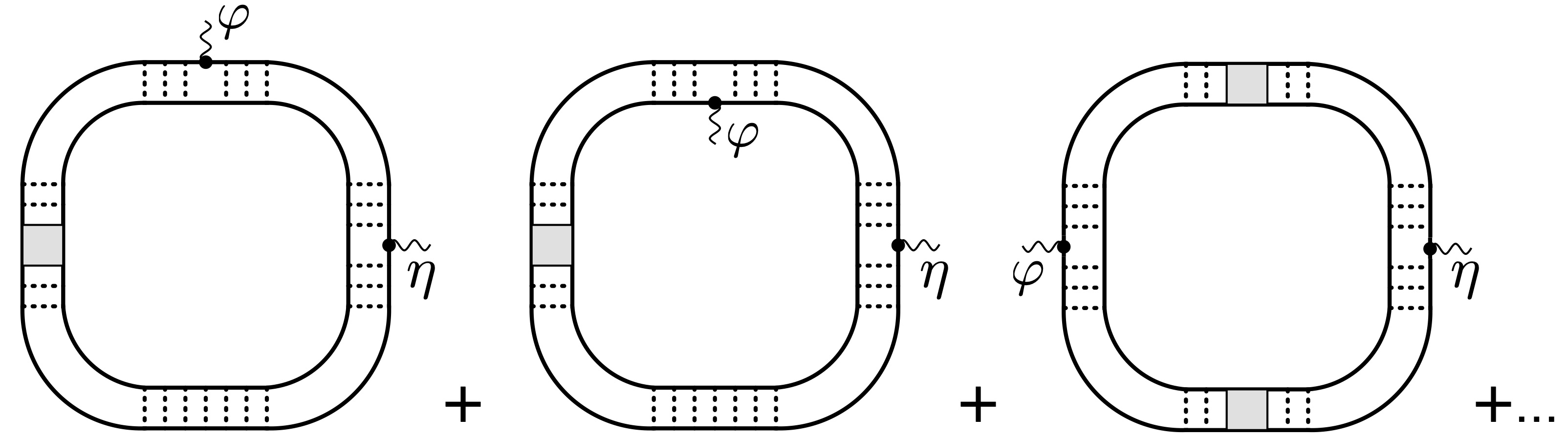}
\caption{Diagrams contributing to $\chi_{kn}^{st,\eps}$.
}
\label{fig:static_eps}
\end{figure}
\begin{figure}[tb]
\includegraphics[width=6cm]{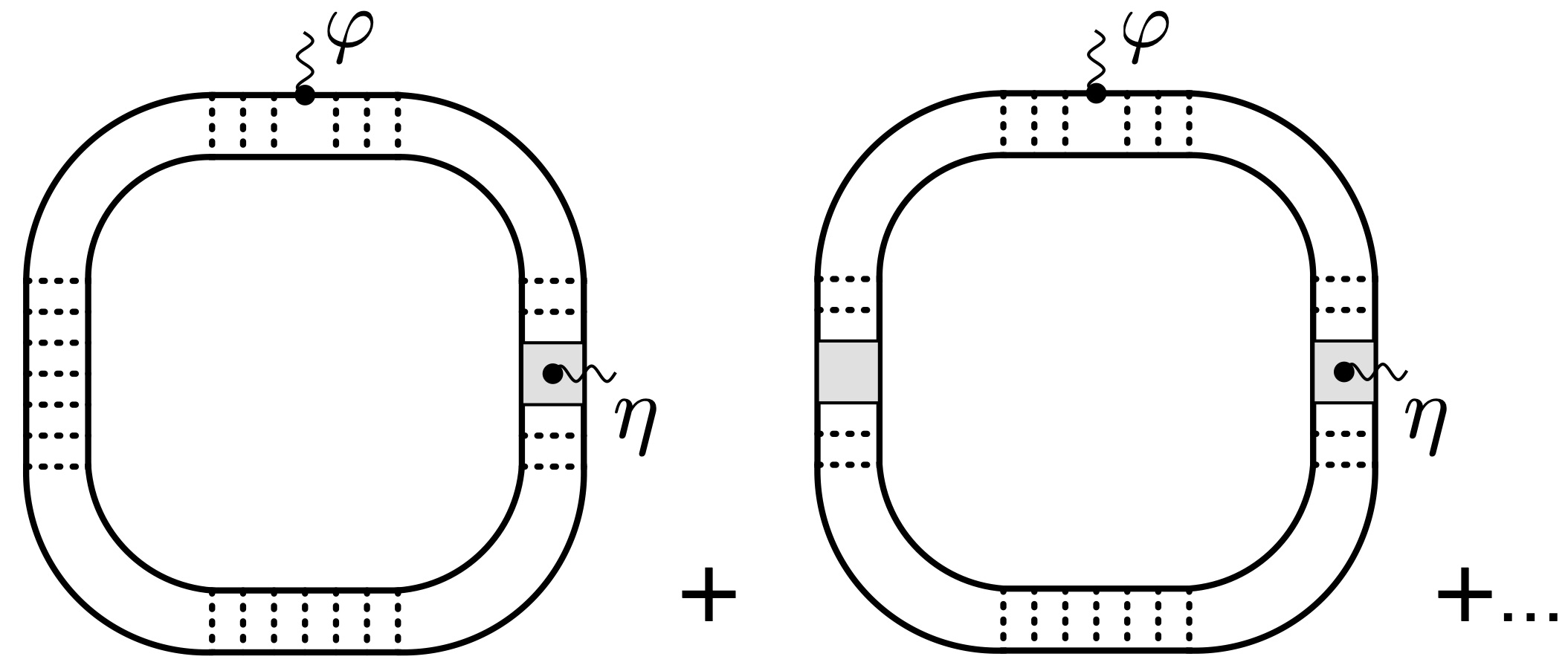}
\caption{Diagrams contributing to $\chi_{kn}^{st,\Gamma}$.}
\label{fig:static_Gamma}
\end{figure}
\begin{figure}[tb]
\includegraphics[width=6cm]{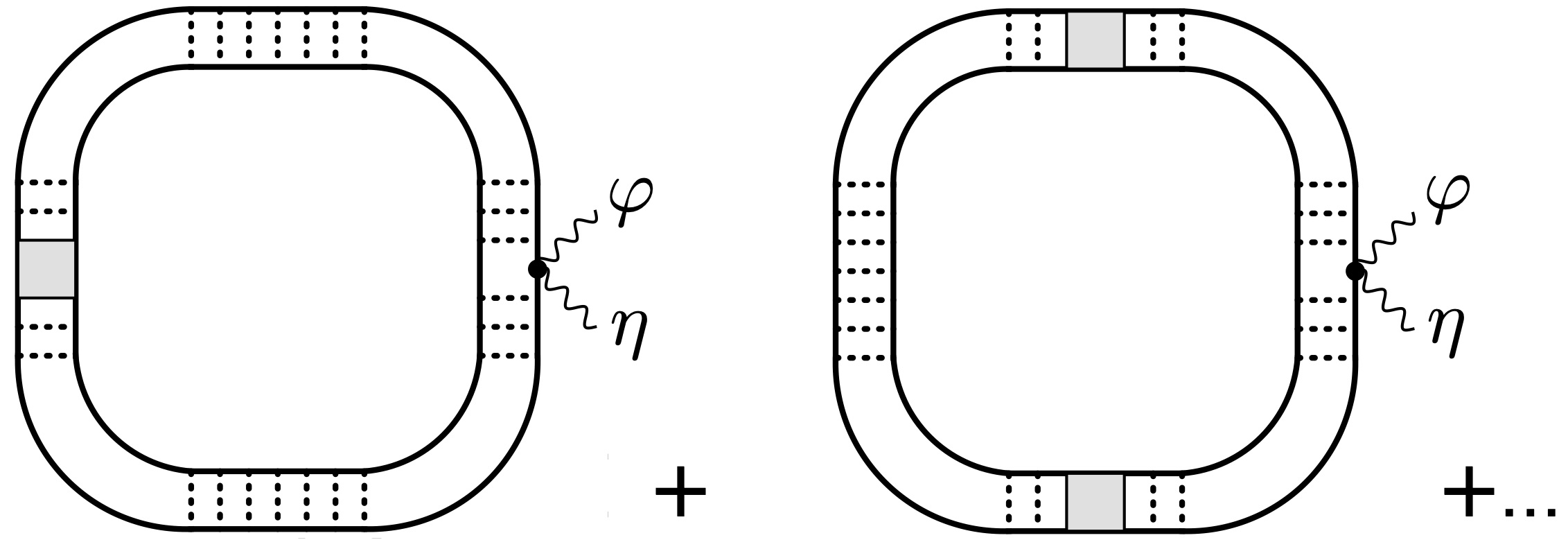}
\caption{Diagrams contributing to $\chi_{kn}^{st,n}$. 
}
\label{fig:static_n}
\end{figure}
We label the contributions to $\chi_{kn}^{st}$ as $\chi_{kn}^{st,\eps}$, $\chi_{kn}^{st,\Gamma}$ and $\chi_{kn}^{st,n}$ and display the corresponding diagrams in Figs.~\ref{fig:static_eps},~\ref{fig:static_Gamma} and \ref{fig:static_n}, respectively. The term $k^{dm}_{\varphi_1,n}$ depends on the potential $\varphi_1$ explicitly. Therefore, averaging can be performed with the $\varphi_1$-independent part of the action. By comparison with Eq.~\eqref{eq:ndmstart} it is immediately clear that $\chi_{kn}^{st,n}=-n^{dm}$. The other two terms, $k^{dm}_{\varphi_1,\eps}$ and $k^{dm}_{\varphi_1,\Gamma}$, have an implicit $\varphi_1$-dependence due to the averaging with respect to the $\varphi_1$-dependent action. We need to expand the expressions for $k^{dm}_{\varphi_1,\eps}$ and $k^{dm}_{\varphi_1,\Gamma}$ and the action up to second order in the generators $\hat{P}$. In order to calculate the contributions arising from the differentiation $\partial_{\varphi_1}$, it is instructive to note that for $\varphi_2=0$, as relevant here, and due to the relation $\tr[\delta \hat{Q}]=0$, the potential $\varphi_1$ enters the action only in the combination $\varphi_1 \tr[(\nabla\hat{P})^2]$. Another useful observation is that at second order in the generators $\hat{P}$ the interaction part of the action becomes independent of the diffusion coefficient, and so does the expression for $k^{dm}_{\varphi,\Gamma}$ before averaging. We can therefore perform the averages for $k^{dm}_{\varphi,\eps}$ and $k^{dm}_{\varphi,\Gamma}$ in Eq.~\eqref{eq:kdmphieps} and Eq.~\eqref{eq:kdmphiGamma} with the help of the action taken at $\varphi_1=0$, if we trade the differentiation with respect to $\varphi_1$ for a differentiation with respect to (minus) the chemical potential $\mu$. The chemical potential, in turn, enters the action only through the diffusion coefficient. This argument allows us to immediately obtain the relation $\chi_{kn}^{st,\eps}+\chi_{kn}^{st,\Gamma}=-\partial_\mu k^{dm}$. As a consequence, the total diffusion mode contribution to the static part of the correlation function reads as
\begin{align}
\chi_{kn}^{st,dm}&\equiv\chi_{kn}^{st,\eps}+\chi_{kn}^{st,\Gamma}+\chi_{kn}^{st,n}=-\partial_\mu k^{dm}-n^{dm}\no\\
&=-T\partial_T n^{dm},
\end{align}
where the last equality was already established in Sec.~\ref{subsec:hd}. We therefore conclude that both routes to the calculation the interaction corrections of $\chi^{st}_{kn}$  presented above give the same result, $\chi_{kn}^{st,dm}=-T\partial_\mu c$, where $c$ is the specific heat. This result is in agreement with Ref.~\cite{Fabrizio91}, where a thermodynamic approach was used.

\section{Derivation of the generalized NL$\sigma$M}
\label{sec:NLSMderivation}

In this section, we present the derivation of the the generalized NL$\sigma$M discussed in Sec.~\ref{sec:genNLSM}. In Sec.~\ref{sec:model}, we introduce the electronic action that serves as a starting point for the derivation. Then, for the sake of clarity, we first focus on the NL$\sigma$M for the non-interacting case in Sec.~\ref{sec:NLSMnoninteracting} before including interactions and source fields in Sec.~\ref{sec:InteractingNLSM}. 

\subsection{Model}
\label{sec:model}

Starting point for our considerations is action 
\be
S_{k}[\psi^\dagger,\psi]=\int_\mathcal{C}dt \int_{\bfr} \left(\psi^\dagger i \partial_t\psi-k[\psi^\dagger,\psi]\right).\label{eq:Kaction}
\ee
This action is defined on the Keldysh time-contour $\mathcal{C}$ \cite{Schwinger61,Kadanoff62,Keldish65,Kamenev11}, which consists of a forward ($+$) and a backward branch ($-$). In Eq.~\eqref{eq:Kaction}, the heat density is defined as $k=h-\mu n$, where $h$, $\mu$, and $n$ are the Hamiltonian density, chemical potential, and particle density, respectively. Further, $\psi=(\psi_\uparrow,\psi_\downarrow)$, $\psi^\dagger=(\psi^*_\uparrow,\psi^*_{\downarrow})$ are Grassmann fields with spin up and spin down components.

The Hamiltonian density $h$ consists of two parts, $h=h_0+h_{int}$, where
\begin{align}
h_{0}(x)&=\frac{1}{2m}\nabla{\psi}^\dagger_x\nabla{\psi}_x+\psi^\dagger_x u_{dis}(\bfr)\psi_x\label{eq:h0},\\
{h}_{int}(x)&=\frac{1}{4 }n(x) \left({F_0^\rho}/{\nu}\right)n(x) +\bfs(x)\left(F_0^\sigma/\nu \right) \bfs(x).\label{eq:hint}
\end{align}
The disorder potential is chosen as delta correlated white noise characterized by $\left\langle u_{dis}(\bfr)u_{dis}(\bfr')\right\rangle=\frac{1}{2\pi\nu \tau}\delta(\bfr-\bfr')$ and $\left\langle u_{dis}(\bfr)\right\rangle=0$. The angular brackets symbolize averaging over different realizations of the disorder potential. We will assume that disorder is weak in the sense that $\eps_F\tau\gg 1$, where $\eps_F$ is the Fermi energy. In Eq.~\eqref{eq:hint}, we introduced the following expressions for the number and spin densities: $n(x)={\psi}^\dagger_x\sigma_0\psi_x$, and $\bfs(x)=\frac{1}{2}{\psi}^\dagger_x \bfsigma \psi_x$. The Pauli matrices $\sigma^l$ for $l\in\{0,1,2,3\}$ act in spin space ($\uparrow,\downarrow$). For the sake of simplicity, we restrict ourselves to a short-range interaction model. Long-range Coulomb interactions can be included into the formalism straightforwardly following Ref.~\cite{Schwiete16a}. 

\subsection{NL$\sigma$M  for non-interacting systems}
\label{sec:NLSMnoninteracting}

Following standard steps in the derivation of the Keldysh NL$\sigma$M, (i) disorder average, (ii) Hubbard Stratonovich transformation of the resulting four-fermion term with the matrix field $\underline{\hat{Q}}=\hat{u}\hat{Q}\hat{u}$, (iii) saddle point approximation $\hat{Q}\rightarrow \hat{\sigma}_3$, and (iv) inclusion of fluctuations $\hat{Q}=\hat{U}\hat{\sigma}_3\hat{\bar{U}}$ with $\hat{U}\hat{\bar{U}}=1$, as described in Appendix \ref{app:derivation}, the electronic part of the action can be presented in the form
\begin{align}
S[\vec{\Psi}^\dagger,\vec{\Psi}]&=\int \vec{\Psi}^\dagger\left(\hat{G}^{-1}+\hat{\bar{U}}[\hat{G}_0^{-1},\hat{U}]\right)\vec{\Psi}.\label{eq:Sforexpansion}
\end{align}
In this equation, we introduced the inverse of the disorder averaged Green's function, $\hat{G}^{-1}=\hat{G}_0^{-1}+\frac{i}{2\tau}\hat{\sigma}_3$, where $\hat{G}_0=\mbox{diag}(G^R_0,G^A_0)$ is a diagonal matrix in Keldysh space, and $G^R_0$ and $G^A_0$ are the non-interacting retarded and advanced Green's functions. The combination $\hat{\bar{U}}[\hat{G}_0^{-1},\hat{U}]$ contains slow gradients of the fields $\hat{U}$ and $\hat{\overline{U}}$, as well as small differences of their frequency arguments. In order to make those explicit, we introduce the fields
\begin{align}
\hat{\mathcal{V}}^i=\hat{\bar{U}}\nabla^i\hat{U}, \qquad  \hat{\mathcal{E}}=\hat{\bar{U}}[\hat{\eps},\hat{U}].
\end{align}
Using this notation, we can write the action as
\begin{align}
&S[\vec{\Psi}^\dagger,\vec{\Psi}]\label{eq:Sslow}\\
&=\int \vec{\Psi}^\dagger\left(\hat{G}^{-1}+ \hat{\mathcal{E}}+\frac{1}{2m}\left[\hat{\mathcal{V}}^i\overrightarrow{\nabla}^i-\overleftarrow{\nabla}^i\hat{\mathcal{V}}^i+\hat{\mathcal{V}}^i\hat{\mathcal{V}}^i\right]\right)\vec{\Psi}.\no
\end{align}
A summation in the vector index $i$ is implied. 

The derivation of the NL$\sigma$M proceeds via an expansion in the slow fields $\hat{\mathcal{E}}$ and $\hat{\mathcal{V}}^i$. The conventional model is obtained by truncating the expansion at the first order in $\hat{\mathcal{E}}$ and at the second order in $\hat{\mathcal{V}}^i$. The resulting sigma model action takes the form $S=\frac{\pi\nu i}{4} \Tr[D(\nabla \hat{Q})^2+4i\hat{\eps} \hat{Q}]$, where $D=v_F^2\tau/d$ is the diffusion coefficient in $d$ dimensions defined with the help of the Fermi velocity $v_F$. From now on we will restrict the discussion to the two-dimensional case $d=2$.  

The expansion in the slow fields requires the evaluation of certain momentum integrals over products of fermionic Green's functions. This integration is simplified by the fact that due to the presence of the slow modes $\hat{U}$ and $\hat{\bar{U}}$ all the fermionic frequencies lie within a small energy shell of order $1/\tau$ around the Fermi surface. In the conventional derivation, it is therefore sufficient to neglect the frequency arguments of the Green's functions entirely. Indeed, a perturbative expansion in those frequencies gives rise to terms that are small in the parameter $\omega/\eps_F$, where $\omega$ is a typical frequency. For our purposes, however, it is crucial to account for such terms, since they encode the particle-hole asymmetry that will render the observables of interest finite. In order to illustrate this point, it is instructive to inspect the term arising at second order in the expansion in $\hat{\mathcal{V}}^i$, 
\begin{align}
\delta S_{\mathcal{V}^2}&=-i\pi \nu \int_{\eps_1,\eps_2,{\bf r}} D_\eps \tr[\hat{\mathcal{V}}^{i\perp}_{\eps_2,\eps_1} \hat{\mathcal{V}}^{i\perp}_{\eps_1,\eps_2}].\label{eq:SV2}
\end{align}
In this formula, we used the notation $\hat{A}^\perp=\frac{1}{2}[\hat{A},\hat{\sigma}_3]\hat{\sigma}_3$, where $\hat{A}$ is a matrix in Keldysh space, and we wrote $D_\eps=D+\frac{\tau}{m}\eps$ with $\eps=(\eps_1+\eps_2)/2$. The second term in the expression for the diffusion coefficient $D_\eps$ is smaller than the leading term by a factor of $\eps/\eps_F$. The gradient term in the conventional sigma model action is obtained by neglecting the $\eps$-dependence of the diffusion coefficient and using the identity $\tr[\hat{\mathcal{V}}^{i\perp} \hat{\mathcal{V}}^{i\perp}]=-\frac{1}{4}\tr[(\nabla \hat{Q})^2]$. 

\begin{figure}
\includegraphics[width=6cm]{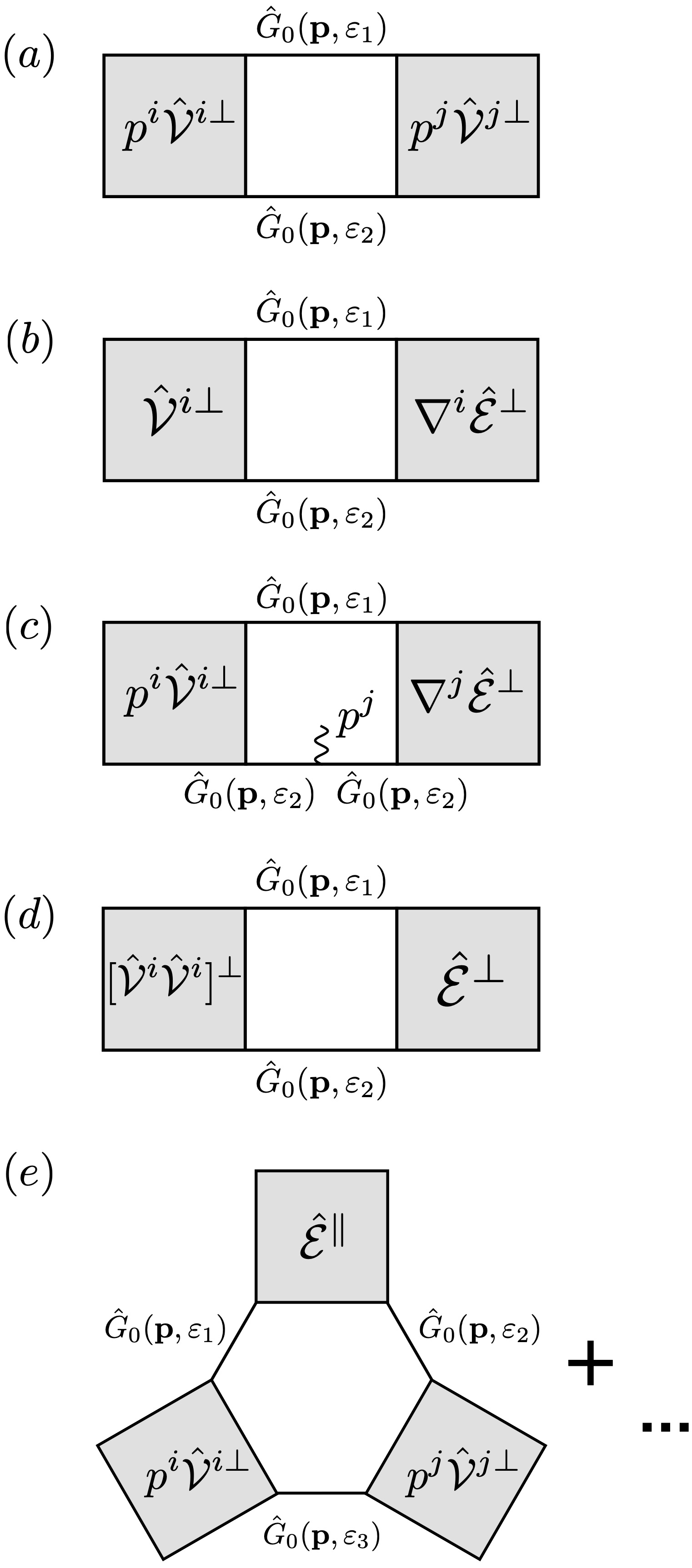}
\caption{Diagrammatic representation of the slow mode expansion resulting in the generalized NL$\sigma$M in Eq.~\eqref{eq:SQ}. The diagrams illustrate traces in Keldysh space. An integration over fast momenta ${\bf p}$ is implied. Slow mode matrices are symbolized by dark squares; the symbol $\perp$ labels matrices that are off-diagonal in Keldysh space, $\parallel$ labels diagonal matrices. The matrix Green's function $\hat{G}_0=(G_0^R,G_0^A)$ contains retarded and advanced Green's functions on its diagonal, so that all traces contain at least one retarded and one advances Green's function. The momentum vertex in (c) arises due to the expansion of $\mathcal{E}$ in coordinates.}
\label{fig:expansion}
\end{figure}

The result \eqref{eq:SV2} is not satisfactory, since it does not allow us to write the action in terms of the field $\hat{Q}$. This shortcoming can be corrected by including additional terms in the slow mode expansion which give contributions of the same order. Relevant contributions are either linear in both $\hat{\mathcal{E}}$ and $\hat{\mathcal{V}}^i$, in which case an additional expansion in coordinates is required, or of second order in $\hat{\mathcal{V}}^i$ and first order in $\hat{\mathcal{E}}$. These terms are illustrated in Fig.~\ref{fig:expansion}. 
The part of the action in Eq.~\eqref{eq:SV2} that accounts for the frequency dependent part of the diffusion coefficient is proportional to $\tr[\hat{\eps} \hat{\mathcal{V}}^{\perp,i} \hat{\mathcal{V}}^{\perp,i}]$. After including the contributions shown in Fig.~\ref{fig:expansion}, this term is replaced by  $\tr[(\hat{\eps}+\hat{\mathcal{E}}) \hat{\mathcal{V}}^{\perp,i} \hat{\mathcal{V}}^{\perp,i}]=-\frac{1}{4} \tr[\hat{\eps}(\nabla \hat{Q})^2]$. After adding the leading contribution to the gradient term in the NL$\sigma$M, which is proportional to $D$ (cf. Eq.~\eqref{eq:SV2}) and the term obtained from the linear expansion in $\hat{\mathcal{E}}$, we arrive at the action
\begin{align}
S_1[\hat{Q}]=&\frac{i\pi\nu}{4}\Tr[\hat{D}_{\hat{\eps}} (\nabla \hat{Q})^2+4i\hat{\eps} \hat{Q}]\label{eq:SQ}
\end{align} 
with $\hat{D}_{\hat{\eps}}=D+\frac{\tau}{m}\hat{\eps}$. The inclusion of the $\varepsilon$-dependent part of the diffusion coefficient encodes the deviations of the single-particle energy from the Fermi energy (and the velocity from the Fermi velocity) in the NL$\sigma$M language. 
In order to complete the derivation of the generalized NL$\sigma$M for non-interacting systems, we also need to include a term with four gradients into the action. Such a term can be obtained in two different ways. The first route amounts to an expansion in the slow modes $\hat{\mathcal{V}}^i$ up to fourth order similar to the expansion in $\hat{\mathcal{V}}^i$ and $\mathcal{E}$ described above (for details see Appendix \ref{sec:S2}, also Ref.~\cite{Wang94}). The resulting term in the action takes the form
\begin{align}
S_{2}[\hat{Q}]&=-\frac{\pi\nu}{8}DD_\eps'\Tr[\nabla^2 \hat{Q}(\nabla \hat{Q})^2].\label{eq:S2}
\end{align}
There is a second contribution to the fourth order term. In the considerations presented so far, we immediately restricted ourselves to the so-called massless manifold described by transversal fluctuations of the form $\hat{Q}=\hat{U}\hat{\sigma}_3\hat{\bar{U}}$. Such fluctuations obey the constraint $\hat{Q}^2=1$. For the discussion of the four-gradient term, however, it important to account for massive (longitudinal) fluctuations around the saddle point $\hat{\sigma}_3$ as well. These fluctuations can mediate a coupling between the soft modes. For many purposes, such an effect can be neglected since the mass $\sim 1/\tau$ of the massive modes makes the resulting contributions to the action small. However, since the particle-hole asymmetric terms are small themselves, the influence of the massive modes can become important. The relevance of this mechanism was first pointed out in Ref.~\cite{Wang94}, see Appendix \ref{app:SM} for details. The result is the additional term $S_M$ displayed in Eq.~\eqref{eq:SMmain}. The fourth-order terms must be treated on the same footing as the contribution proportional to $D_\eps'$ in Eq.~\eqref{eq:SQ}. Indeed, the latter term is smaller by a factor of $D_\eps'\bfq^2$ compared to the frequency term in the conventional NL$\sigma$M (where $q$ is a typical momentum characterizing the diffusion process). Likewise, $S_2$ and $S_M$ are smaller by a the same factor $D_\eps'\bfq^2$ compared to the gradient term in the conventional model.

By combining the three contributions displayed in Eqs.~\eqref{eq:SQ}, \eqref{eq:S2} and \eqref{eq:SMmain} we obtain the generalized NL$\sigma$M for non-interacting systems as $S[\hat{Q}]=S_1[\hat{Q}]+S_2[\hat{Q}]+S_M[\hat{Q}]$. The generalized model can be formulated in two equivalent ways as
\begin{align}
S[\hat{Q}]&=\frac{i\pi\nu}{4}\Tr[D (\nabla \hat{X})^2+4i\hat{\eps} \hat{X}]\\
&+\frac{\pi\nu}{16}DD_\eps'\Tr[\nabla^2 \hat{Q}(\nabla \hat{Q})^2],
\end{align}
or alternatively as
\begin{align}
S[\hat{Q}]&=\frac{i\pi\nu}{4}\Tr[\hat{D}_{\hat{\eps}} (\nabla \hat{Q})^2+4i\hat{\eps} \hat{Q}]\\&-\frac{\pi\nu}{16}DD_\eps'\Tr[\nabla^2 \hat{Q}(\nabla \hat{Q})^2].
\end{align}
This NL$\sigma$M action is the main result of this section. The model is accurate up to leading order in the particle-hole asymmetry terms. The inclusion of the particle-hole asymmetry amounts to the replacement $\hat{Q}\rightarrow \hat{X}$ in the conventional NL$\sigma$M, plus the additional term $S_M$ obtained from the integration of the massive modes.
 
In fact, this innocent-looking extension of the sigma model action has profound consequences. Equipped with the relevant source fields, the model allows for the calculation of quantities such as the thermoelectric transport coefficient that are beyond the reach of the conventional NL$\sigma$M. In this work, we study a model in the unitary symmetry class. An extension of the derivation to, for example, the orthogonal or symplectic symmetry classes is beyond the scope of this work. 

A comment is in order here. The expansion that lead to Eq.~\eqref{eq:SQ} also generates terms that are small in the parameter $\omega\tau$ compared to the leading term in Eq.~\eqref{eq:SV2}, but larger by a factor $\eps_F\tau$ compared to the $\eps$-dependent correction in this equation. Such terms give a contribution of the form $\delta S\sim \nu \tau \Tr[\hat{\eps}\hat{Q} D(\nabla \hat{Q})^2]$ to the action. This is merely a higher order contribution to the gradient expansion of \eqref{eq:Sslow} which does not introduce particle-hole asymmetry. Indeed, one can interpret it as a small correction to the $\tr[\eps \hat{Q}]$-term which shares its symmetries with respect to the frequency structure. The formal symmetry argument presented in Sec.~\ref{sec:pha} leads to the same conclusion. Indeed, $\Tr[\hat{\eps}\hat{Q}D(\nabla\hat{Q})^2]$ is even under the transformation defined in Eq.~\eqref{eq:trafo}. In a similar vein, the expansion up to fourth order in slow gradients also produces a term that is larger than $S_2$ by a factor of $\eps_F\tau$. However, this term contains an even number of $\hat{Q}$ matrices, is even under the transformation \eqref{eq:trafo}, and may therefore be neglected since it does not introduce particle-hole asymmetry into the model.

\subsection{NL$\sigma$M with interactions and source fields}
\label{sec:InteractingNLSM}

We now discuss how the derivation of the NL$\sigma$M presented in Sec.~\ref{sec:NLSMnoninteracting} needs to be modified in order to accommodate interactions and source fields. In order to describe interactions, we add the interaction part ${h}_{int}$ of Eq.~\eqref{eq:hint} to the Hamiltonian density and study ${h}={h}_0+{h}_{int}$. We further include the source terms 
\be
S_\eta&=-2\int_x [\eta_2(x)k_{cl}(x)+\eta_1(x)k_q(x)],\label{eq:Seta}\\
S_\varphi&=-2\int_x [\varphi_2(x)n_{cl}(x)+\varphi_1(x)n_q(x)].\label{eq:Sphi}
\ee
In order to be able to calculate $\chi_{kn}$ and $\chi_{nk}$ in the Keldysh NL$\sigma$M approach, we first define the classical ($cl$) and quantum components ($q$) of the heat density and density symmetrized over the two branches of the Keldysh contour, $k_{cl/q}=\frac{1}{2}(k_+\pm k_-)$, and $n_{cl/q}=\frac{1}{2}(n_+\pm n_-)$, respectively \cite{Kamenev11}. Using these definitions, the retarded correlation functions can be obtained as $\chi_{kn}(x_1,x_2)=-2 i\left\langle k_{cl}(x_1) n_q(x_2)\right\rangle$ and $\chi_{nk}(x_1,x_2)=-2 i\left\langle n_{cl}(x_1) k_q(x_2)\right\rangle$. The angular brackets symbolize averaging is with respect to the action.

The interaction terms in the action resulting from $h_{int}$ can be decoupled with the help of bosonic Hubbard Stratonovich fields $\vartheta_{+}^l$ and $\vartheta_-^l$ on forward and backward paths of the Keldysh contour. Then, after performing the Keldysh rotation [Eq.~\eqref{eq:Keldysh_rotation} in appendix \ref{app:derivation}], the action can be presented in the following form
\be
&&S[\vec{\Psi}^\dagger,\vec{\Psi},\vec{\theta},\hat{\eta}]\label{eq:SKeld}\\
&=&\int_x \;\vec{\Psi}^\dagger \left(i\partial_t-[u_{dis}-\mu](1+\hat{\eta})+\hat{\theta}^l\sigma^l-\hat{\varphi}\right)\vec{\Psi}\no\\
&&-\int_x\frac{1}{2m}\nabla \vec{\Psi}^\dagger(1+\hat{\eta})\nabla\vec{\Psi}+\int_{x}  \;\vec{\theta}^T\frac{\hat{\gamma}_2}{1+\hat{\eta}}f^{-1}\vec{\theta}.\no
\ee
Here, we introduced so-called classical ($cl$) and quantum ($q$) components of the fields $\theta^l_{cl/q}=(\vartheta^l_+\pm\vartheta^l_-)/2$ \cite{Kamenev11}, where $l=0$ stands for the singlet channel, and $l\in\{1,2,3\}$ for three triplet-channel components. The fields $\theta^l_{cl/q}$ are sometimes grouped into an eight-component vector $\vec{\theta}$ with components $\theta_1^l=\theta^l_{cl}$ and $\theta_2^l=\theta_q^l$. The interaction potentials for the singlet and triplet channels are contained in the matrix $f=\mbox{diag}(F_0^\rho,F_0^\sigma,F_0^\sigma,F_0^\sigma)/2\nu$. 

The next steps in the derivation of the NL$\sigma$M are the disorder average, and the decoupling with the matrix field $\underline{\hat{Q}}$. The form of the action in Eq.~\eqref{eq:SKeld} is inconvenient for this purpose, because the gravitational potential enters the disorder term. In order to avoid this complication, we introduce the following transformation  of the fermionic fields, $\vec{\Psi}\rightarrow \sqrt{\hat{\lambda}}\vec{\Psi}$ and $\vec{\Psi}^\dagger\rightarrow \vec{\Psi}^\dagger \sqrt{\hat{\lambda}}$, where $\hat{\lambda}=1/(1+\hat{\eta})$ \cite{Michaeli09,Schwiete14b}. Then, the action can be written as 
\begin{align}
S[\vec{\Psi}^\dagger,\vec{\Psi},\vec{\theta},\vec{\eta}]\label{eq:transformedaction}
&=\frac{1}{2}\int_x \vec{\Psi}^\dagger(i\hat{\lambda}\overrightarrow{\partial}_t-i\overleftarrow{\partial}_t\hat{\lambda})\vec{\Psi}\\
&-\int_x \;\vec{\Psi}^\dagger (\mathcal{O}_{kin}+u_{dis}-\mu-\hat{\lambda}\hat{\theta}^l\sigma^l+\hat{\lambda}\hat{\varphi})\vec{\Psi}\no\\
&+\int_x \vec{\theta}^T(\hat{\gamma}_2\hat{\lambda})f^{-1}\vec{\theta}+S_{\mathcal{J}}.\no
\end{align}
Here, the kinetic energy operator $\mathcal{O}_{kin}=-\nabla^2/2m$ was introduced. The term $S_\mathcal{J}$ accounts for the Jacobian of the transformation of the fermionic fields. It is will not play any role in our considerations and we drop it from now on. For a further discussion of this term we refer to Ref.~\cite{Schwiete14b}.

After these preparations, the generalization of the derivation presented in the previous section does not pose a problem. In view of the transformation \eqref{eq:psirot}, the source fields and the Hubbard Stratonovich fields $\hat{\theta}$ are dressed with matrices $\hat{u}$ as $\underline{\hat{\theta}}_{\eps_1\eps_2}=\hat{u}_{\eps_1}\hat{\theta}(\eps_1-\eps_2)\hat{u}_{\eps_2}$. Then, the gradient expansion in the presence of interactions and source fields requires the following replacement, 
\begin{align}
\mathcal{E}\rightarrow \mathcal{E}'=\mathcal{E}+\hat{\bar{U}}\left[\underline{\hat{\Theta}}^l\sigma^l-\frac{1}{2}\{\underline{\hat{\eta}},\hat{\eps}+\underline{\hat{\Theta}}^l\sigma^l\}\right]\hat{U},
\end{align}
where $\vec{\Theta}^0=\vec{\theta}^0-\vec{\varphi}$ and $\vec{\bfTheta}=\vec{\bftheta}$. The result of this procedure can be written as 
\begin{align}
S^0_{\delta Q}&=\frac{\pi\nu i}{4}\Tr\left[ \hat{D}_{\hat{\eps}^{\eta}_{\Theta}}(\nabla \underline{\hat{Q}})^2+4i \hat{\eps}^\eta_\Theta\underline{\delta \hat{Q}} \right]+\int_x \vec{\theta}^T\hat{\gamma}_2\hat{\lambda}f^{-1}\vec{\theta}\no\\
&+2\nu \int_x \vec{\Theta}^T\hat{\gamma}_2\hat{\lambda}\vec{\Theta}-2k_{0}\int_x\eta_2(x)-2n_0\int_x\varphi_2(x)\no\\
&-\frac{\pi\nu}{16}DD_\eps'\Tr[\nabla^2 \hat{Q}(\nabla \hat{Q})^2].\label{eq:S0dQ}
\end{align}
In this equation, we used the notation $\hat{\eps}_\Theta^\eta=\frac{1}{2}\{\hat{\eps}+ \hat{\Theta}^l\sigma^l,\hat{\lambda}\}$. We consistently kept terms up to first order in $\hat{\eta}$ and in $\hat{\varphi}$. The terms in the second line describe contributions originating from the electronic degrees of freedom without participation of the diffusion modes. They arise from diagrammatic blocks containing only retarded or only advanced Green's functions. In the derivation of $S^0_{\delta Q}$ in Eq.~\eqref{eq:S0dQ} terms containing the derivative of the disorder averaged density of states, $\bar{\nu}'_\eps$, have consistently been neglected. As a consequence, the action does not contain a purely electronic contribution that is linear in both $\vec{\eta}$ and $\vec{\Theta}$. We will discuss the impact of such a term on the correlation function in Sec.~\ref{sec:nonconstant}. 

It is often convenient to present the NL$\sigma$M in a form where the Hubbard-Stratonovich fields $\vec{\theta}$ are integrated out. The relevant contraction rules (in the absence of $\hat{\eta}$) are $\langle \theta^0_{i,\bfr,\omega}\theta^{0}_{j,\bfr',-\omega'}\rangle=\frac{i}{2\nu}(\Gamma_{\rho}^0/2)\gamma_2^{ij}
\delta_{\bfr-\bfr'}2\pi\delta_{\omega-\omega'}$ for the charge degrees of freedom, and $\langle \theta^\alpha_{i,\bfr,\omega}\theta^{\beta}_{j,\bfr',-\omega'}\rangle=\frac{i}{2\nu}
(\Gamma_{\sigma}^0/2)\gamma_2^{ij}\delta_{\bfr-\bfr'}2\pi\delta_{\omega-\omega'}\delta_{\alpha\beta}$ for the spin degrees of freedom. The integration in $\vec{\theta}$ leads to the sigma model action in Eq.~\eqref{eq:NLSMgeneral}.

\section{On the role of a non-constant density of states}
\label{sec:nonconstant}

In this section, we discuss the role of the non-constant disorder averaged density of states in two dimensions. To this end it is instructive to revisit the saddle point equation for the matrix $\hat{Q}$, 
\begin{align}
\hat{Q}_0&=\frac{i}{\pi\nu}\left(\hat{G}_0^{-1}+\frac{i}{2\tau}\hat{Q}_0\right)^{-1}.
\end{align}
With the ansatz $Q_0= ({\tau}/{\tau_\eps})\sigma_3$, where $\tau_\eps$ is real, one obtains the following condition for $\tau_\eps$,  
\begin{align}
1=\frac{1}{2\pi\nu \tau}\int_{\bf p} |G^R_\bullet (\bfp,\eps)|^2,\label{eq:cond}
\end{align}
where $G^R_\bullet(\bfp,\eps)=([G_0^R]^{-1}(\bfp,\eps)+i/(2\tau_\eps))^{-1}$. It is instructive to establish a connection between the right hand side of Eq.~\eqref{eq:cond} and the disorder averaged density of states calculated with the help of $G^R_\bullet$,
\begin{align}
\bar{\nu}_\eps=-\frac{1}{\pi}\int_{\bf p} \Im G_\bullet^R(\bfp,\eps)=\frac{1}{2\pi \tau_\eps}\int_{\bf p} |G^R_\bullet(\bfp,\eps)|^2.\label{eq:dos}
\end{align}
By comparison, one finds the relation $\nu\tau=\bar{\nu}_\eps \tau_\eps$. We see that even in two dimensions, when the density of states $\nu$ of the clean system is constant, $\tau_\eps$ acquires an $\eps$-dependence, and so does the disorder averaged density of states $\bar{\nu}_\eps$. The magnitude of this effect can be estimated from Eq.~\eqref{eq:cond}. After transforming the integration measure as $\int_{\bfp}\rightarrow \nu \int_{-\mu}^\infty d\xi_{\bf p}$, it is important for our purposes not to extend the lower limit of the integration range to $-\infty$. The equation for $\bar{\nu}_{\eps}$ obtained after performing the integral in $\xi_{\bf p}$ can be solved approximately and allows us to estimate $\bar{\nu}'_\eps/\nu\sim (\eps_F^2\tau)^{-1}$. The important point is that $\bar{\nu}'_\eps/\nu$ is smaller than $D_\eps'/D$ by a factor $(\eps_F\tau)^{-1}\ll 1$. As explained in Sec.~\ref{sec:hdcf}, this smallness allows us to justify the constant density of states approximation. 

As is clear from the relation $Q_0= ({\tau}/{\tau_\eps})\sigma_3$, the inclusion of a non-constant density of states for the diffusion modes described by the NL$\sigma$M would require a modification of the constraint $\hat{Q}^2=1$ and thus fundamentally alter the structure of the model. Such a generalization would, for example, be necessary to obtain $\chi_{kn,1}^{dyn,0}$ of Eq.~\eqref{eq:chiprime} instead of $\tilde{\chi}_{kn,1}^{dyn,0}$ in Eq.~\eqref{eq:tildechikn1}. The second contribution discussed in Sec.~\ref{sec:ladder}, $\chi_{kn,2}^{dyn,0}$, and the static part $\chi^{st,0}_{kn}$, arise in a different way. They have their origin in a purely electronic contribution,
\begin{align}
S_{\eta\Theta}=-2T\partial_T n_0\int_x \vec{\eta}^T\hat{\gamma}_2\vec{\Theta}^0.
\end{align}
This term was not included in Eq.~\eqref{eq:S0dQ}, because $T\partial_T n_0\propto  \bar{\nu}'_\eps$. As a consequence, $S^0_{\delta Q}$ in Eq.~\eqref{eq:NLSMgeneral} also acquires additional contributions, $S^0_{\delta Q}\rightarrow S^0_{\delta Q}+S_{\eta\varphi}+S_{\eta X}$, where
\begin{align}
S_{\eta X}&=-\frac{\pi}{2} T\partial_T n_0\Gamma_\rho^0\Tr[\hat{\eta}\delta{\hat{\underline{X}}}],\\
S_{\eta\varphi}&=2T\partial_Tn_0\int_x \vec{\eta}^T\hat{\gamma}_2\vec{\varphi}_{FL}.
\end{align}
The static part of the correlation function is obtained from $S_{\eta\varphi}$ as $\chi^{st,0}_{kn}=-T\partial_T n_0 z_1^0$, and $\chi_{kn,2}^{dyn,0}$ originates straightforwardly from $S_{\eta X}$.

\section{Conclusion}
\label{sec:conclusion}

In this manuscript, we introduced a NL$\sigma$M approach aimed at calculating quantities that are strongly affected by particle-hole asymmetry. We focused on two-dimensional systems with quadratic dispersion, and derived a minimal extension of the Finkel’stein model which accounts for deviations of the electron velocity from the Fermi velocity by including a frequency-dependent diffusion coefficient. The generalized model is obtained from Finkel'stein's model by replacing the $\hat{Q}$-field by $\hat{X}=\hat{Q}+\frac{1}{4i}D_\eps'(\nabla\hat{Q})^2$, and adding the contribution $S_M$ to the four-gradient term obtained from the integration of massive modes. Our considerations in this manuscript were based on the Keldysh NL$\sigma$M. Due to the structural similarity, we expect the replacement rule $Q\rightarrow X$ to be applicable for the NL$\sigma$M in the Matsubara formalism as well. The term $S_M$ is also straightforwardly generalized.

We studied a model with short-range Fermi-liquid interactions. Coulomb interactions can be included into the formalism using the procedure outlined in Ref.~\cite{Schwiete16a}. As an application, we analyzed the heat density-density correlation function in the ladder approximation, and calculated interaction corrections to its static part. These calculations served two purposes. They demonstrated that the results obtained with the help of the generalized NL$\sigma$M are consistent with results previously obtained by different means,  Ref.~\cite{Fabrizio91}. The calculations also constitute a first step in analyzing interaction corrections to the thermoelectric transport coefficient, a problem that will be addressed in a future publication.

%%%%%%%%%%%%%%%%%%%%%%%%%%
\acknowledgments

The author would like to thank Z.~I.~Jitu, K.~Michaeli, and T.~Micklitz for discussions and A.~M.~Finkel'stein for valuable advice and comments on the manuscript. This work was supported by the College of Arts and Sciences at the University of Alabama and the National Science Foundation (NSF) under Grant No. DMR-1742752.

\appendix

\section{Gaussian action}
\label{app:perturbations}

In this appendix, we discuss the Gaussian action resulting from $S_{\delta Q}^0$ [Eq.~\eqref{eq:NLSMgeneral}], and the corresponding contraction rules. In this manuscript, we work with the exponential parameterization,
\be
\hat{U}=\mbox{e}^{-\hat{P}/2},\quad \hat{\overline{U}}=\mbox{e}^{\hat{P}/2},\quad \{\hat{P},\hat{\sigma}_3\}=0\label{eq:U}.
\ee
The matrix $\hat{Q}$ is related to $\hat{P}$ as $\hat{Q}=\hat{\sigma}_3 \exp(\hat{P})$. We further write $\hat{P}$ as a matrix in Keldysh space in the form
\be
\hat{P}_{\eps\eps'}(\bfr)=\left(\ba{cc} 0& d_{cl;\eps\eps'}(\bfr)\\d_{q;\eps\eps'}(\bfr)&0\ea\right),
\ee
where $d_{cl/q}$ are hermitian matrices in the frequency domain and in spin space, $[d^{\alpha\beta}_{cl/q;\eps\eps'}]^*=d^{\beta\alpha}_{cl/q;\eps'\eps}$. Expanding $S_{0,\eta\varphi}+S_{int,\eta}$ up to second order in $\hat{P}$ and neglecting the source fields, one finds the Gaussian action as
\be
&&S=-\frac{i\pi\nu}{4}\int\tr[\hat{D}_{\hat{\epsilon}}(\nabla \hat{P})^2-2i\hat{\eps}\hat{\sigma}_3\hat{P}^2]\label{eq:Psq}\\
&&-\frac{\pi^2\nu}{8}\int_{\bfr,\eps_i} \Big(\tr[\hat{\gamma}_i\underline{\hat{\sigma}_3\hat{P}_{\eps_1\eps_2}}]\hat{\gamma}_2^{ij}\Gamma_\rho^0 \tr[\hat{\gamma}_j \underline{\hat{\sigma}_3\hat{P}_{\eps_3\eps_4}}]\no\\
&&+\tr[\hat{\gamma}_i\bfsigma \underline{\hat{\sigma}_3\hat{P}_{\eps_1\eps_2}}]\hat{\gamma}_2^{ij}\Gamma_\sigma^0\tr[\hat{\gamma}_j\bfsigma \underline{\hat{\sigma}_3\hat{P}_{\eps_3\eps_4}}]\Big)\delta_{\eps_1-\eps_2,\eps_4-\eps_3}.\no
\ee
This result allows us to find Gaussian averages of the components of $d_{cl}$ and $d_{q}$ by inverting the quadratic form. In order to formulate the result, it is convenient to separate the singlet and triplet channels. To this end, we expand $d_{cl}$ and $d_{q}$ in terms of the Pauli spin matrices $\sigma^l$ as
\be
d_{cl/q;\eps\eps'}^{l}=\frac{1}{2}\sum_{\alpha\beta}\sigma_{\beta\alpha}^ld^{\alpha\beta}_{cl/q;\eps\eps'},\qquad l=(0,1-3).
\ee
Using this notation, we obtain for the singlet channel ($l=0$)
\begin{align}
&\left\langle d^{0}_{cl;\eps_1\eps_2}(\bfq) d^0_{q;\eps_3\eps_4}(-\bfq)\right\rangle=-\frac{1}{\pi\nu}\mathcal{D}_{\eps}(\bfq,\omega)\left(\delta_{\eps_1,\eps_4}\delta_{\eps_2,\eps_3}\right.\no\\
&\left.-
\delta_{\omega,\eps_4-\eps_3}i\pi\Delta_{\eps_1\eps_2}
{\Gamma}^0_\rho\mathcal{D}^{-1}(\bfq,\omega)\mathcal{D}_1(\bfq,\omega)\mathcal{D}_{\tilde{\eps}}(\bfq,\omega)\right),\label{eq:dsinglet}
\end{align}
Singlet channel and triplet channels do not interfere in the Gaussian approximation. The average in the triplet channel for $i,j\in\{1,2,3\}$ reads as
\begin{align}
&\left\langle d^{i}_{cl;\eps_1\eps_2}(\bfq) d^j_{q;\eps_3\eps_4}(-\bfq)\right\rangle=-\frac{1}{\pi\nu}\delta^{ij}\mathcal{D}_\eps(\bfq,\omega)\left(\delta_{\eps_1,\eps_4}\delta_{\eps_2,\eps_3}\right.\no \\
&\left.-\delta_{\omega,\eps_4-\eps_3}i\pi\Delta_{\eps_1\eps_2}\Gamma^0_\sigma \mathcal{D}^{-1}(\bfq,\omega)\mathcal{D}_2(\bfq,\omega)\mathcal{D}_{\tilde{\eps}}(\bfq,\omega)\right),\label{eq:dtriplet}
\end{align}
In Eqs.~\eqref{eq:dsinglet} and \eqref{eq:dtriplet}, we used the following notation: $\omega=\eps_1-\eps_2$, $\eps=(\eps_1+\eps_2)/2$, $\tilde{\eps}=(\eps_3+\eps_4)/2$, $\Delta_{\eps,\eps'}=\mathcal{F}_\eps-\mathcal{F}_{\eps'}$ and $\delta_{\eps,\eps'}=2\pi \delta(\eps-\eps')$. Further, in addition to the diffusons already introduced in Sec.~\ref{sec:ladder}, we defined 
\begin{align}
{\mathcal{D}}_2(\bfq,\omega)=\frac{1}{D\bfq^2-iz^0_{2}\omega},\label{eq:Dzone}
\end{align}
where $z^0_2=1-\Gamma^0_\sigma$. When the frequency dependence of $\mathcal{D}_\eps$ is neglected, the contraction rules stated in Eq.~\eqref{eq:dsinglet} and Eq.~\eqref{eq:dtriplet} reduce to those obtained from the Keldysh sigma model in the absence of the particle-hole asymmetry.

\section{Derivation of Eq.~\eqref{eq:Sforexpansion}}
\label{app:derivation}

In this appendix, we briefly summarize the initial steps in the derivation of the NL$\sigma$M which lead to Eq.~\eqref{eq:Sforexpansion}. The derivation of the Keldysh NL$\sigma$M was first described in Refs.~\cite{Horbach93,Chamon99,Kamenev99,Feigelman00}. Here, we will use the notation introduced in Ref.~\cite{Schwiete14}. Starting from the action $S_k$ in Eq.~\eqref{eq:Kaction}, it is convenient to form vectors $\vec{\psi}=(\psi_+,\psi_-)^T$, and $\vec{\psi}^\dagger=(\psi^\dagger_+,\psi^\dagger_-)$ with components corresponding to the fields $\psi$ and $\psi^\dagger$ on the forward and backward path, respectively. Then, the Keldysh rotation can be introduced by transforming the fields $\vec{\psi}$ and $\vec{\psi}^\dagger$ as 
\begin{align}
\vec{\Psi}^\dagger=\vec{\psi}^\dagger \hat{L}^{-1}, \quad \vec{\Psi}=\hat{L}\hat{\sigma}_3, \quad \hat{L}=\frac{1}{\sqrt{2}}\left(\ba{cc} 1&-1\\1&1\ea\right).\label{eq:Keldysh_rotation}
\end{align} 
The disorder average of the Keldysh partition function results in a four-fermion term $S_{dis}=(i/4\pi\nu\tau)\int_{\bfr}(\int_t \vec{\Psi}^\dagger_{x}\vec{\Psi}_{x})^2$ in the action. The term $S_{dis}$ can be decoupled with the help of a Hubbard-Stratonovich transformation by introducing an auxiliary integration over a hermitian matrix $\underline{\hat{Q}}(\bfr,t,t')$. As a result, the following terms appears in the action, $\delta S=\frac{i}{2\tau}\int_{\bfr,t,t'}   \vec{\Psi}^\dagger_{\bfr,t}\underline{\hat{Q}}(\bfr,t,t')\vec{\Psi}_{\bfr,t'}+\frac{\pi\nu i}{4\tau}\int_{\bfr,t,t'} \tr[\underline{\hat{Q}}(\bfr,t,t')\underline{\hat{Q}}(\bfr,t',t)]$. The saddle point equation for $\underline{\hat{Q}}$ can be solved by the matrix $\underline{\hat{Q}_0}(\bfr,t,t')=\hat{\Lambda}_{t-t'}$, where $
\hat{\Lambda}_{\eps}=\hat{u}_\eps\hat{\sigma}_3\hat{u}_\eps$ and $\hat{u}_\eps$ was introduced in Eq.~\eqref{eq:defu}. 

We are interested in describing the slow diffusive motion of electrons at long times and distances. In the NL$\sigma$M formalism, the diffusive behavior is encoded in gapless fluctuations around the saddle point solution $\underline{\hat{Q}_0}$ that respect the condition $(\underline{\hat{Q}}\circ\underline{\hat{Q}})_{t,t'}=\delta(t-t')$. The symbol $\circ$ denotes a convolution in time. A convenient parametrization of the fluctuations reads as
\be
\underline{\hat{Q}}=\hat{u}\circ \hat{Q}\circ \hat{u},\quad  \hat{Q}=\hat{U}\circ\hat{\sigma}_3\circ \hat{\overline{U}},
\ee
where $\hat{U}=\hat{U}_{t,t'}(\bfr)$, and $(\hat{U}\circ \hat{\overline{U}})_{t,t'}=\delta(t-t')$. All these steps are standard in the derivation of the Keldysh NL$\sigma$M.  
We present the fermionic part of the action as
\begin{align}
S[\vec{\Psi}^\dagger,\vec{\Psi}]=\int_{x,x'}\vec{\Psi}^\dagger_x\left(\underline{\hat{G}_0^{-1}}(x,x')+\delta_{\bfr,\bfr'}\frac{i}{2\tau} \underline{\hat{Q}}(\bfr,t,t')\right)\vec{\Psi}_{x'}.
\end{align}
Here, $\underline{\hat{G_0}}$ is the non-interacting Green's function of the clean system with the typical triangular structure
\begin{align}
\underline{\hat{G}_0}=\left(\ba{cc} G^R_0& G^K_0\\ 0& G^A_0\ea\right)=\hat{u}\circ \hat{G}_0\circ\hat{u},
\end{align}
where $\hat{G}_0=\mbox{diag}(G^R_0,G^A_0)$ is diagonal, and $G^R_0$, $G^A_0$, and $G^K_0$ are the retarded, advanced and Keldysh components, respectively. In order to prepare the gradient expansion, we rotate the fields $\vec{\Psi}$ and $\vec{\Psi}^\dagger$ as 
\begin{align}
\vec{\Psi}\rightarrow \hat{u}\circ\hat{U}\circ\vec{\Psi}, \quad \vec{\Psi}^\dagger\rightarrow \vec{\Psi}^\dagger \circ\hat{\bar{U}}\circ \hat{u},\label{eq:psirot}
\end{align} 
respectively. It is convenient to Fourier transform all fields with respect to the time arguments, and to use a matrix notation for the resulting fields in the frequency space. This brings us to Eq.~\eqref{eq:Sforexpansion} in the main text.

\section{Derivation of $S_2$ in Eq.~\eqref{eq:S2}}
\label{sec:S2}

The term $S_2$ displayed in Eq.~\eqref{eq:S2} is obtained from Eq.~\eqref{eq:Sslow} by integrating $\vec{\Psi}^\dagger$ and $\vec{\Psi}$ while retaining terms of fourth order in gradients of the slow field $\hat{U}$ and $\hat{\bar{U}}$. Such terms originate from the following two contributions to the action in Eq.~\eqref{eq:Sslow}, $S_\mathcal{V}=\frac{1}{2m}\int_\bfr \vec{\Psi}^\dagger [\hat{\mathcal{V}}^i\overrightarrow{\nabla}^i-\overleftarrow{\nabla}^i\hat{\mathcal{V}}^i]\vec{\Psi}$ and $\mathcal{S}_{\mathcal{V}^2}=\frac{1}{2m}\int_\bfr \vec{\Psi}^\dagger\hat{\mathcal{V}}^i\hat{\mathcal{V}}^i\vec{\Psi}$, upon averaging with $S_0= \int_\bfr \vec{\Psi}^\dagger \hat{G}^{-1}\vec{\Psi}$. Six terms can give rise to four gradients, $\delta S_a=-\frac{i}{4!}\langle\!\langle S_{\mathcal{V}}^4\rangle\!\rangle$, $\delta S_b=-\frac{1}{2!}\langle\!\langle S^2_\mathcal{V} S_{\mathcal{V}^2}\rangle\!\rangle$, $\delta S_c=\frac{i}{2}\langle\!\langle S_{\mathcal{V}^2}^2\rangle\!\rangle$, $\delta S_d=\frac{i}{2}\langle\!\langle\mathcal{S}^2_\mathcal{V}\rangle\!\rangle$, $\delta S_e=i\langle \!\langle S_\mathcal{V} S_{\mathcal{V}^2}\rangle\!\rangle$, and $\delta S_f=-\frac{1}{3!}\langle\!\langle S_\mathcal{V}^3\rangle\!\rangle$. Here, the double brackets $\langle\!\langle \dots\rangle\!\rangle$ denotes the connected average with respect to $S_0$. The calculation is simplified by the fact that the frequency dependence of the Green's function $\hat{G}$ can be neglected. Furthermore, only terms that are odd under the transformation \eqref{eq:trafo} are relevant for our discussion since such terms reflect the particle-hole asymmetry in the system. We find that the important contributions come from
\begin{align}
\delta S_a=&-8\pi\nu DD_\eps'\langle n^in^jn^kn^l\rangle \Tr[\hat{\mathcal{V}}^{i\perp}\hat{\mathcal{V}}^{j\perp} \hat{\mathcal{V}}^{k\parallel} \hat{\mathcal{V}}^{l\parallel}\hat{\sigma}_3],\no\\
\delta S_b=&-\pi\nu DD_\eps'\times\no\\
&\times \Tr[\hat{\mathcal{V}}^i\hat{\mathcal{V}}^i(\hat{\mathcal{V}}^{j\parallel}\hat{\mathcal{V}}^{j\perp}-\hat{\mathcal{V}}^{j\perp}\hat{\mathcal{V}}^{j\perp}-\hat{\mathcal{V}}^{j\perp}\hat{\mathcal{V}}^{j\parallel})\hat{\sigma}_3],\no\\
\delta S_e=&\pi\nu DD_\eps'\Tr[\nabla^i\hat{\mathcal{V}}^{i\perp}(\hat{\mathcal{V}}^j\hat{\mathcal{V}}^j)^\perp\hat{\sigma}_3],\no\\
\delta S_f=&-\pi\nu DD_\eps'\Tr[\mathcal{V}^{k\parallel}[\nabla^j\hat{\mathcal{V}}^{k\perp},\hat{\mathcal{V}}^{j\perp}]\hat{\sigma}_3].
\end{align}
Here, $\langle n^i n^jn^k n^l\rangle=\frac{1}{8}(\delta_{ij}\delta_{kl}+\delta_{ik}\delta_{jl}+\delta_{il}\delta_{jk})$ is an angular average over components of the unit vector ${\bf n}$, summation over repeated vector indices is implied, and we denote the diagonal and off-diagonal components of $\hat{\mathcal{V}}^i$ in Keldysh space as $\hat{\mathcal{V}}^{i\parallel}$ and $\hat{\mathcal{V}}^{i\perp}$, respectively. Summing up all contributions we obtain
\begin{align}
S_2=2\pi\nu DD_\eps'\Tr[\hat{\mathcal{V}}^{i\perp}\hat{\mathcal{V}}^{i\perp}\hat{\mathcal{V}}^{j\perp}\hat{\mathcal{V}}^{j\perp}\hat{\sigma}_3],
\end{align}
which can be brought to the form displayed in Eq.~\eqref{eq:S2} with the help of the identity $\nabla^i\hat{Q}=2\hat{U}\hat{\mathcal{V}}^i\hat{\sigma}_3\hat{\bar{U}}$ in conjunction with the normalization condition $\hat{Q}^2=1$.

\section{Derivation of $S_M$ in Eq.~\eqref{eq:SMmain}}
\label{app:SM}
In this appendix, we describe the derivation of $S_M$ displayed in Eq.~\eqref{eq:SMmain}. In order to understand its origin, it is sufficient to focus on the non-interacting case as described in Sec.~\ref{sec:NLSMnoninteracting}. Retracing the steps outlined in Appendix \ref{app:derivation}, we now generalize the parametrization of the matrix $\hat{Q}$, Eq. (B2), to include massive fluctuations $\hat{Q}\rightarrow \hat{Q}_M=\hat{U} \hat{P}_M \hat{\bar{U}}$ \cite{Pruisken82}. Here, $\hat{P}_M$ is a Hermitian matrix that is block-diagonal in Keldysh space and $\delta \hat{P}_M=\hat{P}_M-\hat{\sigma}_3$ parametrizes massive fluctuations around the saddle point. Correspondingly, the Keldysh partition function is written as $Z=\int_{\Psi^\dagger,\Psi,\hat{P}_M,\hat{U}}I[\hat{P}_M]\exp(iS)$ with
\begin{align}
S=\int \vec{\Psi}\Big(\hat{G}_0^{-1}+\frac{i}{2\tau}\hat{P}_M+\hat{\bar{U}}[\hat{G}_{0}^{-1},\hat{U}]\Big)\vec{\Psi}+\frac{i\pi \nu}{4\tau}\Tr[\hat{P}_M^2].
\end{align}
In the expression for the partition function, $I[\hat{P}_M]$ is the Jacobian arising due the parametrization of $\hat{Q}_M$. With the definition $\hat{G}_{M}^{-1}=\hat{G}^{-1}+\frac{i}{2\tau}\delta\hat{P}_M$, the partition function after integration over the fermionic fields can be presented as  $Z=\int_{\hat{P}_M,\hat{U}}\mbox{e}^{iS}$ with $S=S[\hat{U},\delta \hat{P}_M]+S[\delta \hat{P}_M]$ and
\begin{align}
S[\hat{U},\delta \hat{P}_M]=&-i\tr\ln[1+\hat{G}_{M}\hat{\bar{U}}[\hat{G}_{0}^{-1},\hat{U}]],\label{eq:SUdP}\\
S[\delta \hat{P}_M]=&-i\tr\ln\left[1+\hat{G}\frac{i}{2\tau}\delta \hat{P}_M\right]+\frac{i\pi\nu}{4\tau}\tr[\hat{P}_M^2]\no\\
&-i\ln I[ \hat{P}_M].\label{eq:SdP}
\end{align}
Here, $S[\hat{U},\delta \hat{P}_M]$ describes the coupling of soft and massive modes. 

The influence of the massive modes was entirely neglected in the expansion described in Sec.~\ref{sec:NLSMnoninteracting}, which was based on $S[\hat{U},\delta \hat{P}_M=0]$. This expansion led to $S_1$, Eq.~\eqref{eq:SQ}, and $S_2$, Eq.~\eqref{eq:S2} (which equals $S_{0,\eta\varphi}^{(2)}$ in the notation of Sec.~\ref{subsec:NLSMaction}). The integration of the massive modes can produce a contribution to the NL$\sigma$M action with four gradients, $S_M$ [Eq.~\eqref{eq:SMmain}], of the same form as $S_2$ obtained from $S[\hat{U},\delta \hat{P}_M=0]$. To obtain this term, it is sufficient to integrate $\delta \hat{P}_M$ in the Gaussian approximation. Therefore, $S[\delta \hat{P}_M]$ should be expanded up to second order in $\delta \hat{P}_M$. Upon substituting $\hat{P}_M=\sigma_3+\delta \hat{P}_M$, linear terms in $\delta \hat{P}_M$ cancel between the first two terms in Eq.~\eqref{eq:SdP} by virtue of the saddle point approximation. Higher order terms in $\delta \hat{P}_M$ resulting from the expansion of the $\tr\ln$ in Eq.~\eqref{eq:SdP} give subleading contributions (in the parameter $1/\eps_F\tau$), since they come with an integration in $\xi_{\bfp}$ over a product of only retarded (or only advanced) Green's functions. The Jacobian $I[\hat{P}_M]$ is not easily evaluated in a continuum model, as it requires a regularization. However, from diagrammatic considerations one expects deviations from the self-consistent Born approximation underlying the saddle point equation to be suppressed by powers of $(\eps_F \tau)^{-1}$. In effect, we approximate the quadratic form in $\delta \hat{P}_M$ by $S[\delta \hat{P}_M]\approx \frac{i\pi\nu}{4\tau}\tr[\delta \hat{P}_M^2]$.

\begin{figure}[t!]
\includegraphics[width=4.5cm]{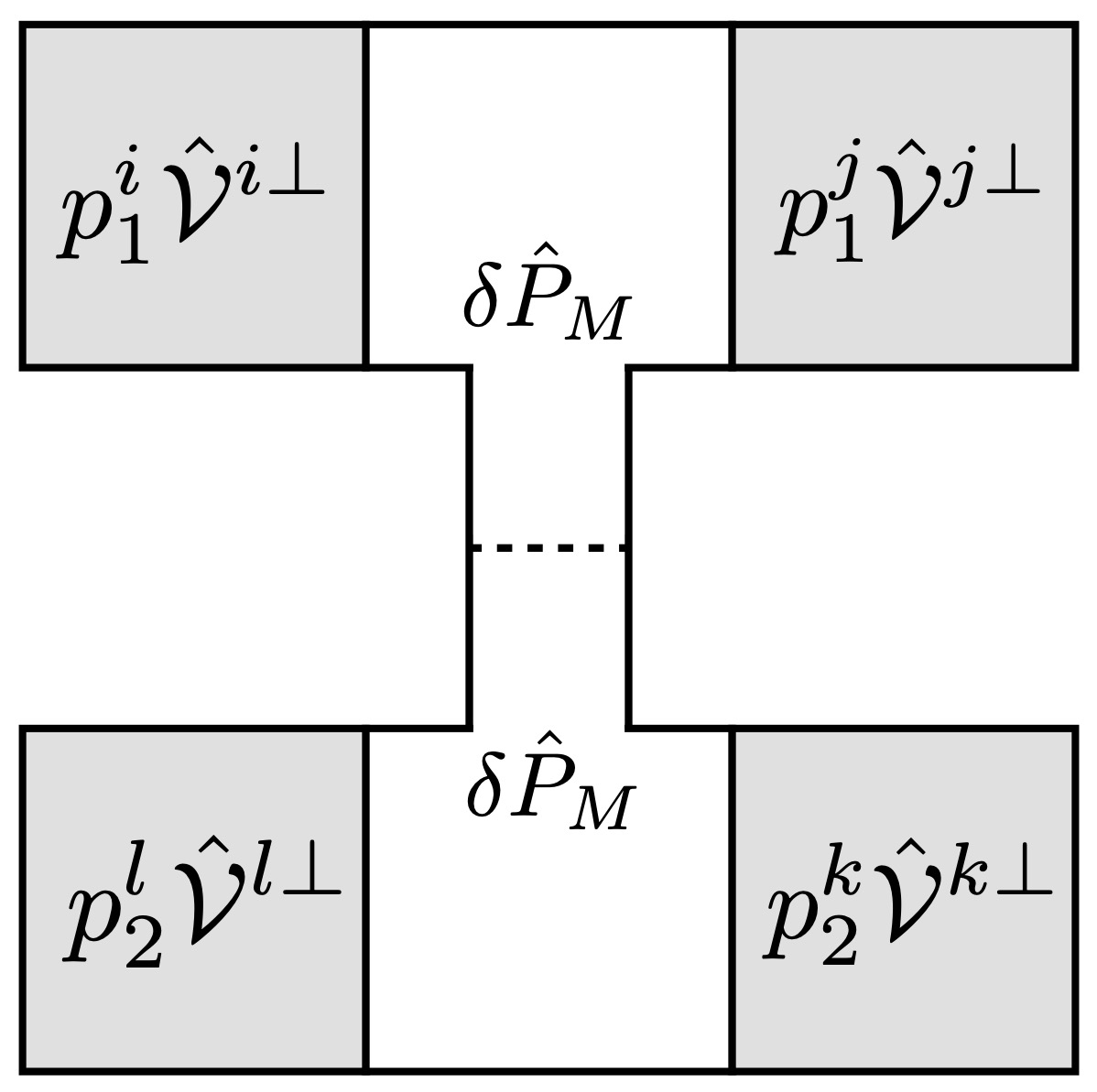}
\caption{Generation of the four-fermion term $S_M$ through the coupling of soft and massive modes.}
\label{fig:massive}
\end{figure}
Corrections to the NL$\sigma$S originating from the coupling of soft and massive modes in $S[\hat{U},\delta \hat{P}_M]$ can be organized as a cumulant expansion in $\delta S=S[\hat{U},\delta \hat{P}_M]-S[\hat{U},\delta \hat{P}_M=0]$. $\delta S$, in turn, is obtained by expanding $G_M$ in powers of $\delta \hat{P}_M$. At first order, the cumulant expansion gives $\delta S^{(1)}=\langle \delta S\rangle$ (where $\langle \dots \rangle$ stands for a Gaussian average with the action $S[\delta \hat{P}_M]$). Such terms can be checked to give small corrections only. The contribution of interest originates from the second cumulant $S_M=\frac{i}{2}\langle\!\langle (\delta S)^2\rangle\!\rangle$. Specifically, the relevant part of $\delta S$ is found by replacing $\hat{\bar{U}}[\hat{G}_{0}^{-1},\hat{U}]\rightarrow \mathcal{O}=\frac{1}{2m}[\hat{\mathcal{V}}^i\overrightarrow{\nabla}^i-\overleftarrow{\nabla}^i\hat{\mathcal{V}}^i]$ in Eq.~\eqref{eq:SUdP}, expanding the logarithm to second order in $\mathcal{O}$, and further expanding one of the two Green's functions in the resulting expression for $\delta S$ to first order in $\delta \hat{P}_M$ as $\hat{G}_M\approx \hat{G}-\frac{i}{2\tau} \hat{G}\delta \hat{P}_M\hat{G}$. After averaging with respect to $\delta \hat{P}_M$, one finds
\begin{align}
S_M=\frac{i}{4\pi\nu\tau}\int d\bfr \;\tr[ (\hat{G}\mathcal{O}\hat{G}\mathcal{O}\hat{G})^\parallel_{\bfr,\bfr} (\hat{G}\mathcal{O} \hat{G}\mathcal{O}\hat{G})^\parallel_{\bfr,\bfr}].
\end{align}
Fig.~\ref{fig:massive} displays the corresponding diagram. Focusing only on the particle-hole asymmetric contribution, one obtains 
\begin{align}
S_M=-\pi\nu DD'\Tr[\sigma_3\hat{\mathcal{V}}^{i\perp}\hat{\mathcal{V}}^{i\perp}\hat{\mathcal{V}}^{j\perp}\hat{\mathcal{V}}^{j\perp}],
\end{align}
which results in Eq.~\eqref{eq:SMmain}.
The described mechanism for producing the four-gradient term through the coupling of soft and massive modes was already noticed in Ref.~\cite{Wang94}. For a comparison with this work, it is instructive to note the relation $\Tr[\nabla^2 \hat{Q}(\nabla \hat{Q})^2]=-\Tr[(\nabla\hat{Q})^2(\nabla\hat{Q})^2 \hat{Q}]$.


\begin{thebibliography}{49}
\bibitem[{Tra()}]{Transposition}
\bibinfo{note}{Note that the transposition $T$ comprises a sign change of the
  momentum operator}.
\end{thebibliography}
\end{document}